\documentclass[fleqn,usenatbib]{mnras}

\usepackage[T1]{fontenc}
\usepackage{ae,aecompl}

\usepackage{graphicx}	
\usepackage{amsmath}	
\usepackage{amssymb}	

\usepackage[dvipsnames]{xcolor}
\hypersetup{
  colorlinks   = true, 
  urlcolor     = RoyalBlue, 
  linkcolor    = RoyalBlue, 
  citecolor    = MidnightBlue 
}
\usepackage{natbib}
\usepackage{color}
\usepackage{rotating}
\usepackage{url}
\usepackage{ifthen}

\usepackage{bm}
\usepackage{bbold}
\usepackage{aas_macros}
\usepackage{verbatim}
\usepackage{algorithm}
\usepackage{algpseudocode}
\usepackage{enumitem}
\usepackage{subfig}

\usepackage{tablefootnote}
\usepackage[flushleft]{threeparttable}

\usepackage{newtxtext, newtxmath}

\DeclareMathAlphabet{\mathpzc}{OT1}{pzc}{m}{it}

\newcommand{\mvec}[1]{\bm{#1}}
\newcommand{\mmat}[1]{\mathbf{#1}}

\newcommand{\myvec}[1]{\mvec{#1}}
\newcommand{\mymat}[1]{\mmat{#1}}
\newcommand\Ztilde{\stackrel{\sim}{\smash{\mymat{Z}}\rule{0pt}{1.1ex}}}
\newcommand\Ztildee{\stackrel{\sim}{\smash{\mymat{Z}}\rule{0pt}{0.8ex}}}
\newcommand{\argmax}[1]{\underset{#1}{\operatorname{argmax}}\,}
\newcommand{\argmin}[1]{\underset{#1}{\operatorname{argmin}}\,}
\newcommand\numberthis{\addtocounter{equation}{1}\tag{\theequation}}

\SetSymbolFont{symbols}{bold}{OMS}{cmsy}{b}{n}
\DeclareSymbolFont{bmisymbols}{OML}{cmm}{b}{it}

\title[AI-driven spatio-temporal engine for lensed SNe]{AI-driven spatio-temporal engine for finding gravitationally lensed type Ia supernovae}

\author[D. Kodi Ramanah, N. Arendse, R. Wojtak]{Doogesh Kodi Ramanah,$^{1}$\thanks{ramanah@nbi.ku.dk} Nikki Arendse,$^{1,2}$ Rados\l{}aw Wojtak$^{1}$ \\
$^{1}$ DARK, Niels Bohr Institute, University of Copenhagen, Jagtvej 128, 2200 Copenhagen, Denmark \\
$^{2}$ Oskar Klein Centre, Department of Physics, Stockholm University, SE-106 91 Stockholm, Sweden
}

\date{Accepted XXX. Received YYY; in original form ZZZ}

\pubyear{2022}

\begin{document}
\label{firstpage}
\pagerange{\pageref{firstpage}--\pageref{lastpage}}
\maketitle

\begin{abstract}
We present a spatio-temporal AI framework that concurrently exploits both the spatial and time-variable features of gravitationally lensed supernovae in optical images to ultimately aid in future discoveries of such exotic transients in wide-field surveys. Our spatio-temporal engine is designed using recurrent convolutional layers, while drawing from recent advances in variational inference to quantify approximate Bayesian uncertainties via a confidence score. Using simulated Young Supernova Experiment (YSE) images of lensed and non-lensed supernovae as a showcase, we find that the use of time-series images adds relevant information from time variability of spatial light distribution of partially blended images of lensed supernova, yielding a substantial gain of around 20 per cent in classification accuracy over single-epoch observations. Preliminary application of our network to mock observations from the Legacy Survey of Space and Time (LSST) results in detections with accuracy reaching around 99 percent. Our innovative deep learning machinery is versatile and can be employed to search for any class of sources which exhibit variability both in flux and spatial distribution of light.
\end{abstract}




\section{Introduction}
\label{intro}

Strong gravitational lensing of astrophysical transient sources, such as supernovae and active galactic nuclei (AGN), involves bending of light from the background source as the light travels towards the observer through a lens galaxy, producing multiple images of the strongly lensed object. Time delays between these images are particularly sensitive to the cosmic expansion rate and observations of several gravitationally lensed quasars were successfully used to place competitive constraints on the Hubble constant \citep{wong2020holicow,2020Birrer}. Time delay distances combined with observations of type Ia supernovae were also used to measure curvature of the Universe \citep{Collett2019, arendse2019lowredshift}.

Type Ia supernovae, as considered in this work, possess several advantages over AGN, such as quasars, as time delay indicators. Since they are standardisable candles, they can be used to directly compute the lensing magnification, thereby mitigating issues posed by the degeneracy between the lens potential and $H_0$ \citep{oguri2003gravitational}, although severe limitations arise from stochastic magnifications due to microlensing \citep{dobbler2006finite, yahalomi2017quadruply}. With exceptionally well-characterised light curve morphology, extracting the time delays is less complicated relative to quasars which display significant variation in their spectral sequences \citep{nugent2002Kcorrections, pereira2013spectrophotometric, rodney2016Refsdal}. Finally, after the supernova has faded away, follow-up observations may be conducted to better constrain the lens galaxy properties, such as the mass profile, without contamination from the supernova \citep{ding2021improved}. However, the expected time delays from type Ia supernovae are typically much shorter and image separations are smaller than in known lensed quasars, making the precise measurement of time delays more challenging.

Lensed supernovae remain, nevertheless, exotic astrophysical objects, with the challenges inherent to their discovery rendering them extremely difficult to observe. The two main indicators are the multiply-imaged lensing signature ({\it image multiplicity}) \citep{oguri2010gravitationally} and exceptionally amplified ({\it magnification}) flux \citep{goldstein2017glSNe}. As of date, only three multiply-imaged resolved systems have been discovered \citep{kelly2015multiple, goobar2017iptf16geu, rodney2021gravitationally}, with two of them lensed by galaxy clusters and one by a galaxy lens, along with a handful of highly magnified unresolved detections \citep{amanullah2011highly, quimby2014detection, rodney2015illuminating}, all discovered serendipitously rather than by means of dedicated searches. A major impediment is that these supernovae are visible for at most hundred days after explosion, in contrast to the much longer timescales of quasars. Consequently, it is of paramount importance to flag lensed candidates while the supernova is still active so that high-resolution imaging or spectroscopy may be used to confirm the lensed nature and to measure a time delay, thereby further limiting the window of opportunity. The situation is yet exacerbated due to most strong gravitational lenses yielding image separations that are below the resolution of ground-based optical surveys \citep{oguri2006image}.

Given the scarcity of gravitationally lensed astrophysical supernovae and the typical volume of imaging data sets, machine learning (ML) techniques, adept at detecting subtle patterns and correlations in data, provide a natural solution to this problem. As such, various ML algorithms \citep{lanusse2018deeplens, schaefer2018deep, davies2019using, avestruz2019automated, cheng2020identifying,  huang2020finding, canameras2020holismokesII, huang2020discovering, gentile2022lenses} for finding strongly lensed systems, particularly well-suited for hosting lensed supernovae or quasars, in photometric surveys have been recently developed. These ML studies typically employ deep convolutional neural networks that are tuned to identify the characteristic signature of the arc features constituting the Einstein rings.

In this work, we aim to harness the power of artificial intelligence (AI), in particular, ML algorithms, to design a new tool specifically tailored for finding gravitationally lensed type Ia supernovae from the Young Supernova Experiment \citep{jones2020YSE}. Our approach introduces some innovative features with respect to the existing ML-based lens finders to cope with the distinct challenges posed by the Young Supernova Experiment, as outlined below. Our proposed methodology is, nevertheless, also of practical relevance for other current or next-generation surveys, as long as the images constituting the training set accurately emulates the characteristics of the actual observations.

\begin{figure}
	\centering
		{\includegraphics[width=\hsize,clip=true]{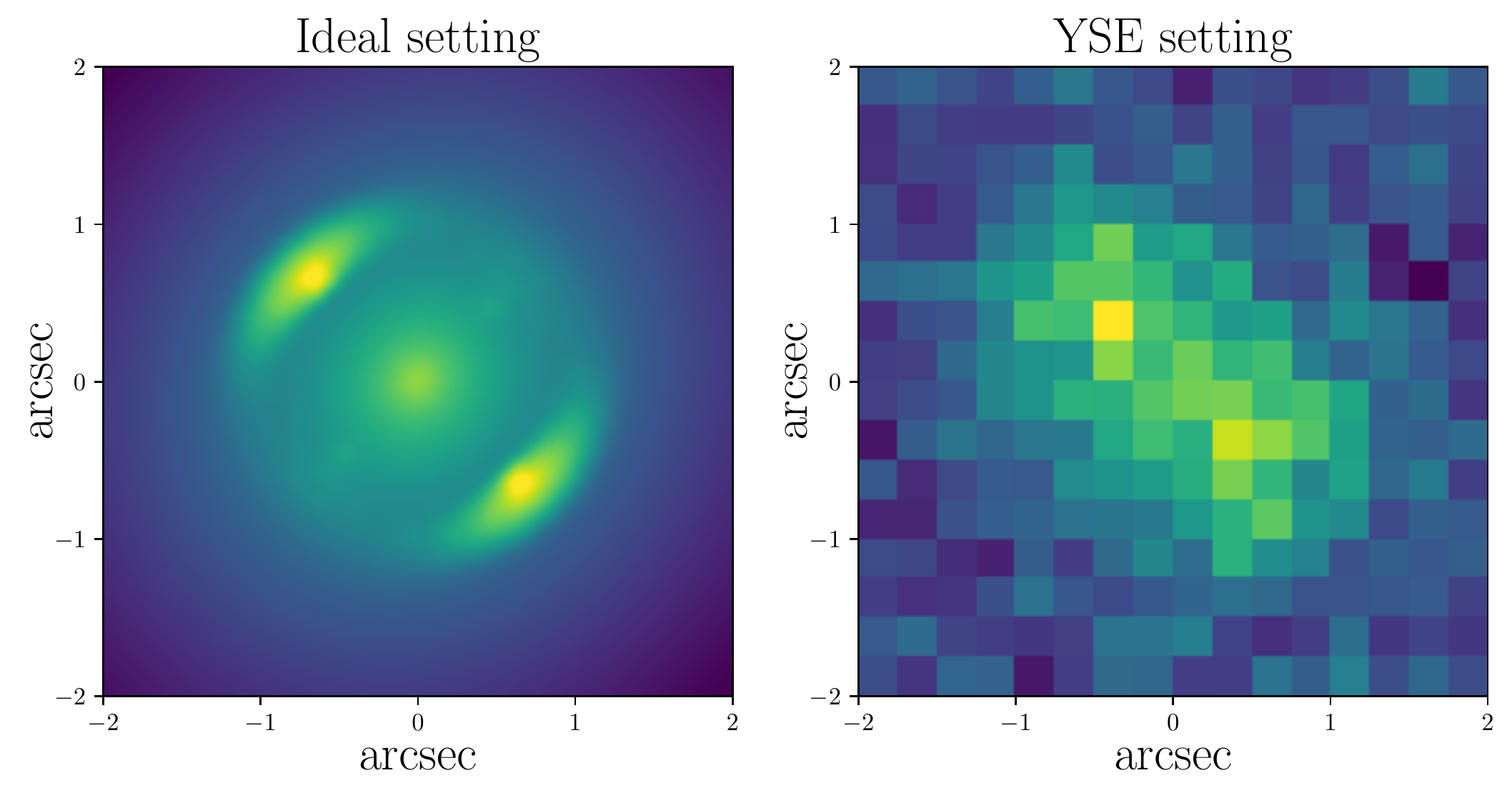}}
	\caption{Contrast between a simulated gravitationally lensed supernova as imaged by a perfect instrument ({\it left panel}) and the Young Supernova Experiment (YSE, {\it right panel}), highlighting the challenging nature of discovering lensed supernovae from YSE observations. The differences between the idealised and YSE settings lie mainly in the resolution, background noise level and quality of seeing, as characterised by the instrument's PSF.}
	\label{fig:ideal_vs_YSE_comparison}
\end{figure}

\begin{figure}
	\centering
		{\includegraphics[width=\hsize,clip=true]{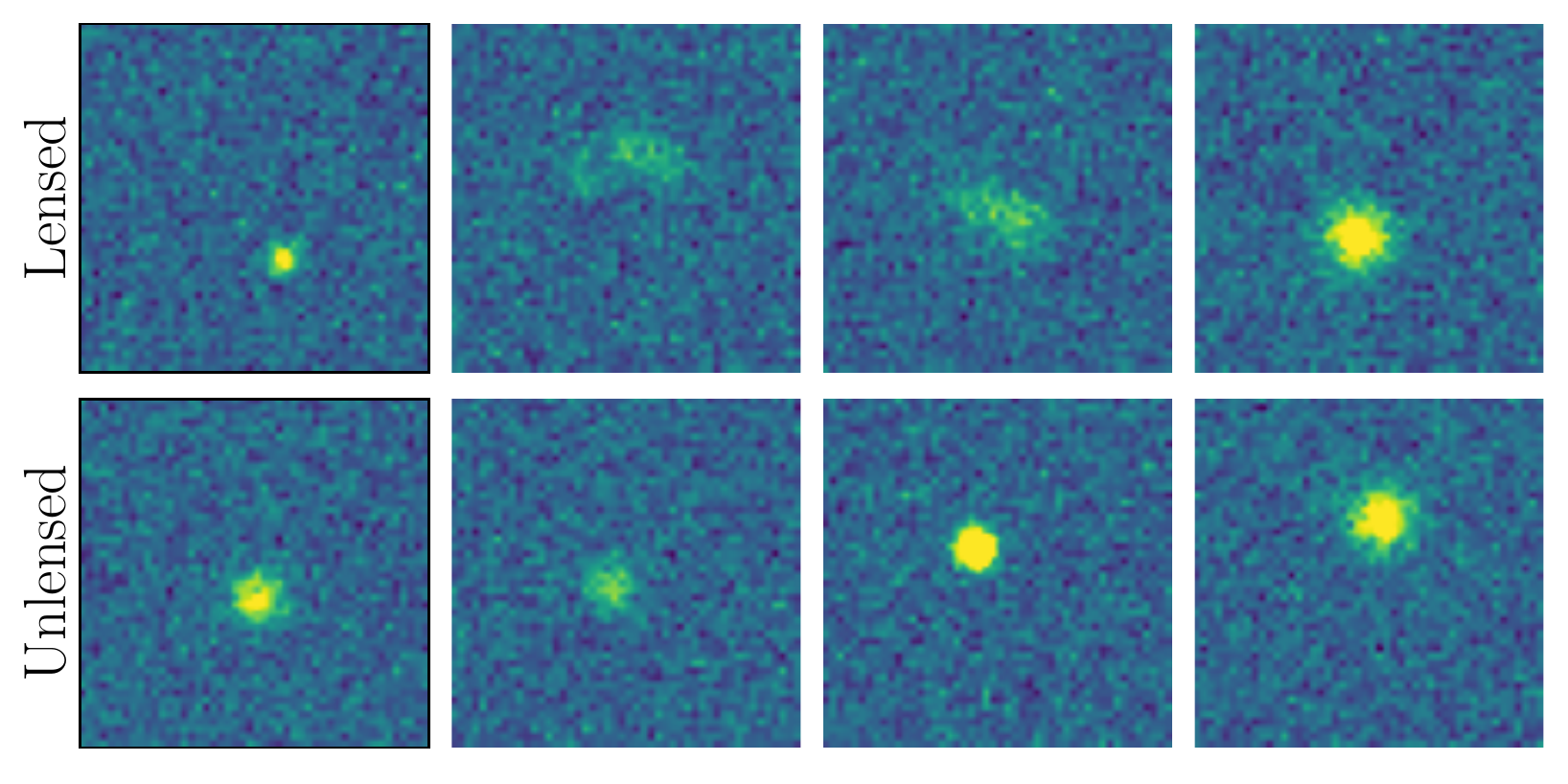}}
	\caption{A random set of lensed ({\it top row}) and unlensed ({\it bottom row}) supernovae simulated according to YSE $i$-band specifications. The colour scale is anchored for all images, with each image stamp extending over 12$^{\prime \prime}$. Our image simulation pipeline is designed such that the properties of the mock unlensed images closely match those of the actual YSE observations. These two rows illustrate that most lensed sources in YSE will be unresolved, thereby appearing similar to unlensed ones. Since the YSE data pipeline provides difference images of transients directly, i.e. the subtraction of a historic reference image from a newly observed image, we do not incorporate the lensing galaxy in our image simulation pipeline.}
	\label{fig:ensemble_images}
\end{figure}

The Young Supernova Experiment \citep{jones2020YSE} (hereafter YSE) constitutes a recent endeavour of a novel optical time-domain survey on the Pan-STARRS telescopes. YSE is geared towards the discovery of fast-rising supernovae within a few hours to days of explosion, thereby complementing other currently ongoing surveys. Moreover, YSE is presently the only time-domain survey with observations in four bands, with the capacity to discover faint transients ($\sim 21.5$~mag in {\it gri} and $\sim 20.5$~mag in {\it z}) that allow the study of the earliest phases of stellar explosions. According to survey forecasts from simulations, the full-capacity operation of YSE will result in the discovery of around 5000 new supernovae per year, with at least a couple of them within three days of explosion per month. The latter YSE attribute is especially pertinent as it is essential to identify new lensed supernova candidates sufficiently early so as to initiate the follow-up sequence in a timely fashion. The substantial volume of new observations from YSE is, therefore, an exciting hunting ground for lensed supernovae. Monte Carlo simulations of strongly lensed supernovae with observationally motivated supernova rates and properties of lens galaxies show that YSE has the potential to discover two lensed supernovae per year \citep{wojtak2019magnified}.

Such an undertaking is, however, riddled with challenges in the YSE context. It is extremely unlikely that YSE will fully resolve the particularly conspicuous signature of the Einstein rings in lensed systems. This is mainly due to the instrumental capabilities of the Pan-STARRS telescopes, such as the point spread function (PSF) and the typical background noise level. As an illustration of the innate difficulty of this task, Fig.~\ref{fig:ideal_vs_YSE_comparison} depicts the stark contrast between a gravitationally lensed supernova in an ideal setting with almost perfect observing conditions, i.e. remarkably high resolution, very sharp PSF and non-existent background noise, and the same source as observed in typical YSE settings. Furthermore, unresolved or partially resolved lensed images are a common feature for YSE-like observations. An additional limitation emanates from the fact that for typical lensing configurations predicted for the survey, the second brightest image has flux near or below the detection limit \citep{wojtak2019magnified}. As a consequence, the aforementioned ML-based lens finders are not immediately adequate for our particularly strenuous problem due to the absence of the distinctive features of such lensed systems. Fig.~\ref{fig:ensemble_images}, displaying a random sample of simulated lensed and unlensed (i.e. normal) supernovae further corroborates the challenging nature of finding lensed supernovae from the YSE survey.

\begin{figure*}
	\centering
		{\includegraphics[width=\hsize,clip=true]{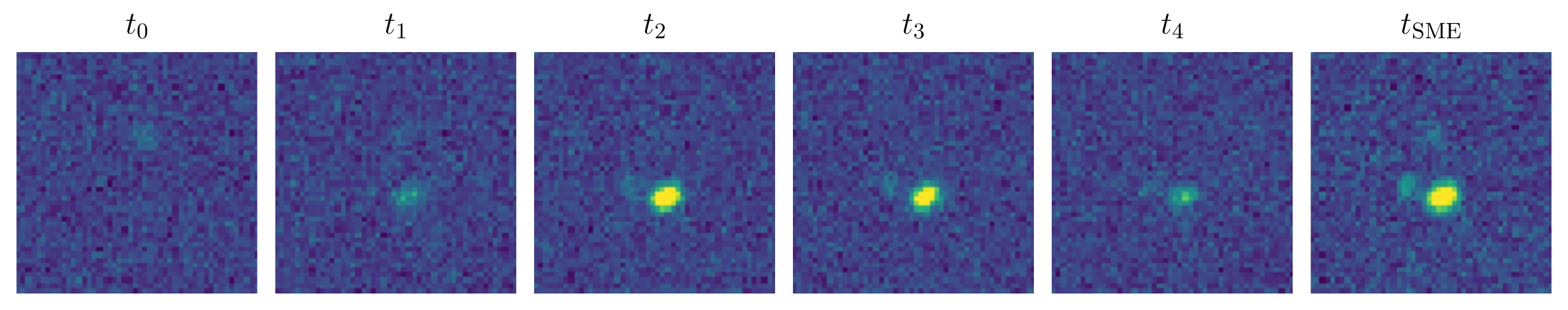}}
	\caption{Temporal evolution of the image stamp, as observed by YSE, of a simulated realisation of a lensed supernovae. The image on the far right corresponds to the compressed time-series representation via smooth manifold extraction (SME).}
	\label{fig:temporal_evolution_images}
\end{figure*}

To make tangible progress in the search for lensed supernovae in the YSE context, there is a pressing need for highly effective detection algorithms. To this end, we design a neural network that is sensitive to both spatial and temporal correlations and takes into account the temporal evolution of an astrophysical transient over multiple epochs, in addition to its spatial information from a single-epoch image. An example of an extra distinctive trait, even for unresolved images, is a spatial shift in the light centroid of the lensed point sources over sequential epochs, contrary to unlensed sources. Searches for sources with spatially extended and time-variable sources were also proposed as a method to select lensed quasar candidates \citep{Kochanek2006, Lacki2009, Chao2020}. This constitutes the rationale behind our novel approach where we employ time-series images in training a spatio-temporal engine, based on recurrent convolutional layers, to perform a binary classification of lensed and unlensed type Ia supernovae. We build on recent advances pertaining to variational inference to yield approximate Bayesian uncertainties on the neural network predictions. A measure of the network confidence associated with a given prediction yields an additional metric to aid in swiftly pinpointing promising lensed candidates that can be prioritised during the subsequent human vetting step. The above innovative aspects distinguish our approach from previous ML-based lens finders proposed in the literature.

To illustrate the improvement in classification accuracy, by virtue of the informative features encoded in the respective neural network inputs, we design and implement three separate neural classifiers:
\begin{enumerate}[label=(\Alph*),leftmargin=*]
    \setlength\itemsep{0.01em}
    \item Single-epoch model trained using single-epoch images randomly selected from the time-series images;
    \item Compressed temporal model trained using compressed 2D temporal representations of the time-series images;
    \item Spatio-temporal model trained using all time-series images.
\end{enumerate}
All above ML models output a network probability score, but are fed distinct types of input images. The time-series images may be unevenly distributed in time. To obtain the compressed temporal representation, we implement a variant of the smooth manifold extraction (SME) algorithm \citep{shihavuddin2017SME}, recently proposed in the field of medical image processing, to combine the arbitrary number of time-series images per source into a single informative image.

An illustration of the temporal evolution of the observed image for a particular simulated lensed supernova is provided in Fig.~\ref{fig:temporal_evolution_images}, along with the compressed SME representation. Nevertheless, the SME compression inevitably leads to some loss of temporal information. In contrast, the spatio-temporal model fully exploits the temporal correlations between the multiple-epoch images and can handle arbitrary number of time-series images per source. For the single-epoch model, lensing can manifest itself solely via spatial extension of the source due to partial blending of lensed images, while for the spatio-temporal one, both spatial and temporal dependence of the flux are used to differentiate between lensed and unlensed supernovae. The latter model is also more sensitive to the distinctive patterns of the variation in brightness of lensed sources compared to the compressed temporal network.

The remainder of this paper proceeds as follows. Section~\ref{methods} describes the relevant details pertaining to the background, numerical implementation and optimisation of our ML and image simulation algorithms. We then demonstrate the performance of our spatio-temporal model in Section~\ref{results}, followed by a summary of the main findings and potential extensions of our study in Section~\ref{conclusions}.

\section{METHODS}
\label{methods}

This section covers the design and implementation of the Bayesian neural classifiers, the generation of simulated lensed and unlensed YSE-like time-series images constituting our training, validation and test sets, and the time-series image compression algorithm.

\subsection{Image simulation procedure}
\label{image_sim_section}

We employ the multi-purpose lens modelling package \textsc{lenstronomy} \footnote{\url{https://lenstronomy.readthedocs.io/en/latest/}} \citep{birrer2018lenstronomy} to generate images that accurately reflect the actual YSE observations. The YSE survey is designed to acquire well-sampled light curves in four bands ({\it griz}) for several thousands of transient astrophysical sources up to a redshift of around 0.2. The current field of view of YSE is approximately 750 square degrees of the sky with a cadence of three days, with the area covered by the survey to be doubled in the near future, with a median seeing (FWHM of PSF) of $1.28^{\prime \prime}$. The image properties and YSE-like survey characteristics, as adopted in this work, are detailed in Table~\ref{tab:YSE_specs}. YSE utilises the {\it difference imaging} technique for the identification of new sources in optical images, whereby a historic reference image is subtracted from a given input image to excise the static and non-varying sources, such that transient sources manifest as residual flux that can be detected and measured photometrically with conventional methods.

\begin{table}
\begin{center}
\renewcommand{\arraystretch}{1.2}
\begin{tabular}{ l l l l}
\hline
{\bf Image property} & {\bf YSE} & {\bf LSST} \\
\hline\hline
PSF $({\rm FWHM})$ & $\mathcal{N}^{\rm \mathcal{S}}(5, 0.95, 0.375)$ & $\mathcal{N}^{\rm \mathcal{S}}(4, 0.55, 0.299)$ \\
Pixel size $(^{\prime \prime})$ & $0.25$ & $0.2$ \\
Pixel background noise ($\sigma_{\mathrm{bkg}}$) & $6.5$ & $6.1$ \\
Exposure time $({\rm s})$ & $27$ & $30$ \\
Image stamp size $(^{\prime \prime})$ & $12.0$ & $9.6$ \\
Image stamp area $({\rm arcsec}^2)$ & $144$ & $92$ \\
Number of pixels & $48 \times 48$ & $48 \times 48$ \\
Cadence (days) & $6$ & $9$ \\
Zero-point magnitude & $24.74$ & $27.79$ \\
Limiting magnitude & $21.40$ & $23.90$ \\
\hline
\end{tabular}
\end{center}
\caption{\label{tab:YSE_specs} \textbf{Simulated image characteristics.} The images constituting the training, validation and test sets are generated in accordance with the YSE survey specifications, thereby closely emulating real YSE observations. The image properties for our preliminary LSST set-up are also provided. For YSE and LSST, the cadence, zero-point and limiting magnitudes correspond to their $i$-band specifications. $\mathcal{N}^{\mathcal{S}}(a, \mu, \sigma)$ denotes a skew-normal distribution with skewness parameter $a$.}
\end{table}

\begin{table}
\begin{center}
\renewcommand{\arraystretch}{1.2}
\begin{tabular}{ l l}
\hline
{\bf Parameter} & {\bf Distribution} \\
\hline\hline
Lens redshift & $z_{\rm lens} \sim \mathcal{N}(0.4, 0.1)$ \\
Lensed source redshift & $z_{\rm src} \sim \mathcal{N}^{\rm \mathcal{S}}(2.5, 0.67, 0.1)$ \\
Unlensed source redshift & $z_{\rm src} \sim \mathcal{N}^{\rm \mathcal{S}}(5, 0.08, 0.1)$ \\
Lensed source position $(^{\prime \prime})$ & $\alpha, \delta \sim \mathcal{U}(-\theta_{\mathrm{E}}, \theta_{\mathrm{E}})$ \\
Unlensed source position $(^{\prime \prime})$ & $\alpha, \delta \sim \mathcal{N}(0, 0.65)$ \\
\hline
{\bf Lens galaxy} \\
\hline \hline
{Elliptical power-law mass} \\
{Lens centre $(^{\prime \prime})$} & $x_\textrm{lens}, y_\textrm{lens} \equiv (0, 0)$ \\
Einstein radius $(^{\prime \prime})$ & $\theta_{\mathrm{E}} \sim \mathcal{N}^{\rm \mathcal{S}}(5.45, 0.14, 0.63)$ \\
Power-law slope & $\gamma_{\rm lens} \sim \mathcal{N}(2.0, 0.1)$ \\ 
Axis ratio & $q_{\rm lens} \sim \mathcal{N}(0.7, 0.15)$ \\ 
Orientation angle (rad) & $\phi_{\rm lens} \sim \mathcal{U}(-\pi/2, \pi/2)$ \\ 
\hline 
{\bf Environment} \\
\hline \hline
External shear modulus & $\gamma_{\rm ext} \ \sim \mathcal{U}(0, 0.05)$ \\
Orientation angle (rad) & $\phi_{\rm ext} \sim \mathcal{U}(-\pi/2, \pi/2) $ \\
\hline
{\bf Light curve} \\
\hline \hline
Stretch & $x_1 \ \sim \mathcal{N}^{\rm \mathcal{S}}(-3.6, 0.96, 1.2)$ \\
Colour & $c \ \sim \mathcal{N}^{\rm \mathcal{S}}(5.5, -0.11, 0.13)$ \\
Absolute magnitude & $M_{\mathrm{abs}} \ \sim \mathcal{N}(-19.43, 0.12)$\\
Milky Way extinction & $E(B-V) \ \sim \mathcal{U}(0, 0.1)$ \\
\hline
\end{tabular}
\end{center}
\caption{\label{tab:param_distributions} \textbf{Parameter distributions for lensed and unlensed systems.} The distribution of input parameters employed in the image simulation pipeline to generate the training, validation, and test data sets. $\mathcal{N}(\mu, \sigma)$ corresponds a normal distribution with mean $\mu$ and standard deviation $\sigma$, $\mathcal{N}^{\mathcal{S}}(a, \mu, \sigma)$ implies a skew-normal distribution with skewness parameter $a$, and $\mathcal{U}(x, y)$ denotes a uniform distribution with bounds $x$ and $y$.}
\end{table}

We consider a supernova as a point source in its host galaxy. Since YSE provides difference images that contain only the astrophysical transients, we simulate images that consist of solely the lensed or unlensed supernova, i.e. the host and lens galaxies are not represented. As such, their respective light profiles are not of relevance to our image simulation pipeline. We assume a standard $\Lambda$CDM cosmological model, as characterised by the latest best-fit values from the Planck Collaboration \citep{planck2018cosmo}. While the actual YSE image stamp extends over 75$^{\prime \prime}$ in length, we consider only the central portion covering an area of 144 arcsec$^2$, thereby obviating image artefacts, such as strips of dead pixels, in the real observations. This cutout size is sufficiently large to encode all relevant lensing features.

A key component in our image generation pipeline is the choice of a model profile with the prerequisite flexibility to sufficiently characterise lensing systems in practice. Previous ML-based studies involving lens modelling made use of the singular isothermal ellipsoid (SIE) lens mass profile \citep{hezaveh2017fast, perrault2017uncertainties, pearson2019use, madireddy2019modular, maresca2021auto, schuldt2021efficient}. An extension to this model is the power-law elliptical lens mass distribution (PEMD) \citep{kormann1994isothermal, barkana1998fast}, where the 3D power-law mass slope $\gamma_{\rm lens}$ is allowed to vary. The PEMD model has been adopted in recent studies \citep{wagnercarena2021hierarchical, park2021largescale, pearson2021strong}, and we employ the same model in our work. The PEMD profile can be expressed in terms of six parameters as
\begin{equation}
    \kappa(x, y) = \frac{3 - \gamma_{\rm lens}}{2} \left(\frac{\theta_{\mathrm{E}}}{\sqrt{q_\textrm{lens} x^2 + y^2/q_\textrm{lens}}} \right)^{\gamma_{\rm lens} - 1} ,
    \label{eq:pemd}
\end{equation}
where $q_\textrm{lens}$ denotes the projected axis ratio of the lens, $\gamma_{\rm lens}$ is the logarithmic slope of the 3D density profile, and $\theta_{\mathrm{E}}$ corresponds to the Einstein radius. The coordinates ($x, y$), by definition, result from a rotation of the sky coordinates by the lens orientation angle $\phi_\textrm{lens}$, such that the $x$-axis is aligned with the major axis of the lens, and are subsequently centred at the position ($x_{\rm lens}, y_{\rm lens}$) of the lens centre. We also account for the external shear component that is characterised by the shear modulus $\gamma_{\rm ext}$ and the shear angle $\phi_{\rm ext}$. To circumvent issues arising from cyclic boundary conditions due to the 2$\pi$-periodic property of the angles, the target lens mass ellipticity and external shear are expressed as follows:
\begin{align}
& e_1 = \frac{1-q_{\text{lens}}}{1+q_{\text{lens}}} \cos (2 \phi_\text{lens}) \\
& e_2 = \frac{1-q_{\text{lens}}}{1+q_{\text{lens}}} \sin (2 \phi_\text{lens}) \\
& \gamma_1 = \gamma_\text{ext} \cos (2 \phi_\text{ext}) \\
& \gamma_2 = \gamma_\text{ext} \sin (2 \phi_\text{ext}) .
\label{eq:shear_conversion}
\end{align}

\begin{figure}
	\centering
		{\includegraphics[width=0.45\textwidth,clip=true]{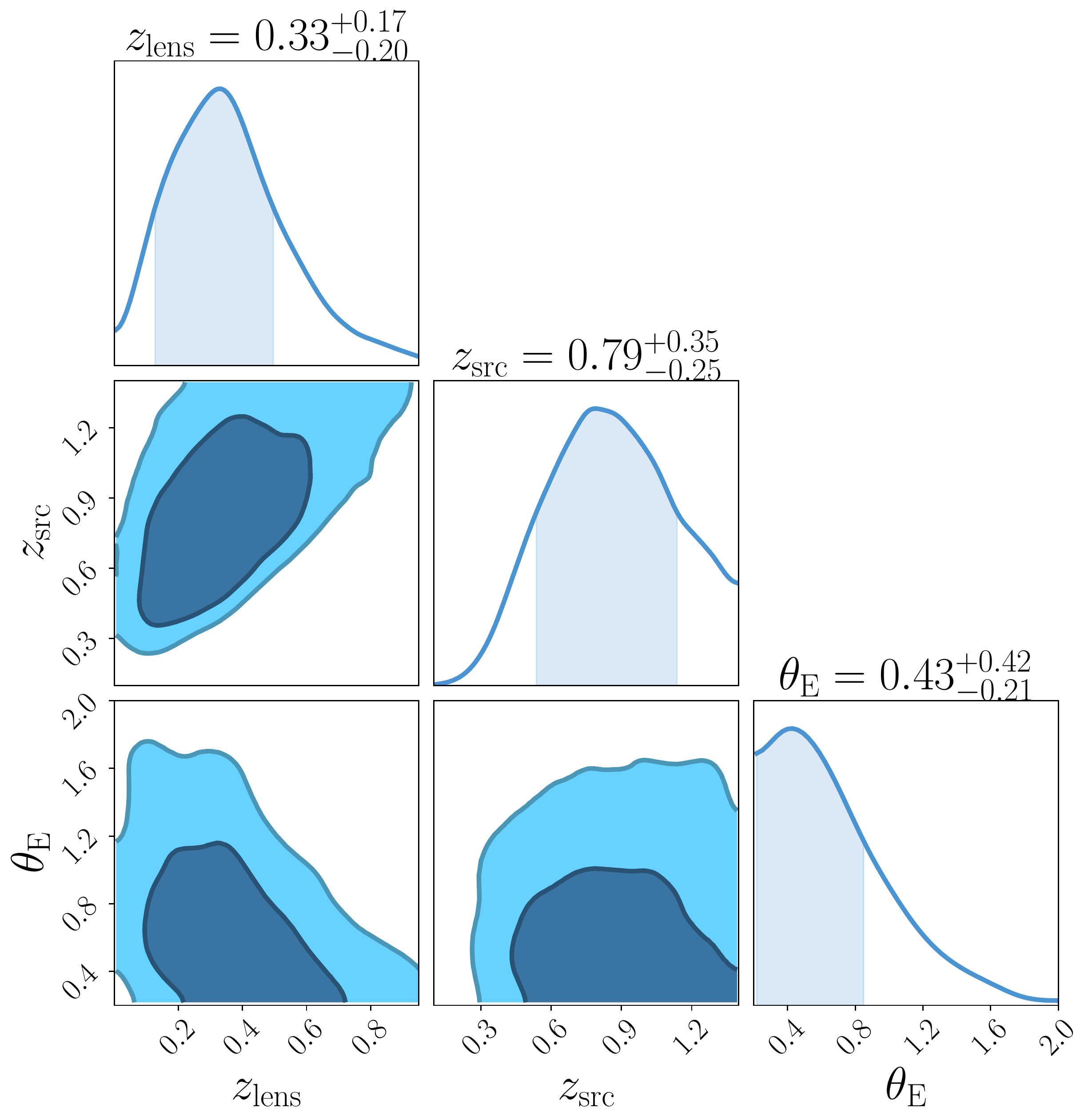}}
	\caption{Joint distribution of the lens ($z_{\textrm{lens}}$) and source ($z_{\textrm{src}}$) redshifts, and Einstein radius ($\theta_{\textrm{E}}$), employed in the image simulation pipeline of lensed supernova systems. The combinations of $\{z_{\textrm{lens}}, z_{\textrm{src}}, \theta_{\textrm{E}}\}$ correspond to galaxy-source configurations that result in strong gravitational lensing, thereby characterising all observable lensed systems given the limiting magnitude of YSE.}
	\label{fig:zl_zs_thetaE_distribution}
\end{figure}

To generate an ensemble of simulated YSE-like images, we sample the relevant parameters from their respective distributions, as detailed in Table~\ref{tab:param_distributions}, motivated by observational and theoretical considerations via Monte Carlo simulations \citep{oguri2010gravitationally, wojtak2019magnified}. Specifically, we employ a joint probability distribution of Einstein radii, lens galaxy redshifts and supernova redshifts for strongly lensed type Ia supernovae detectable in YSE. Using this probability distribution as a prior, we generate lensing configurations by drawing random positions of the source within the area of strong lensing, while keeping the position of the lens fixed to the centre of the image. The joint distribution of redshifts of the lens and source, and Einstein radii, for configurations that induce strong lensing is illustrated in Fig.~\ref{fig:zl_zs_thetaE_distribution}.

For every lensing configuration, we then verify whether the total magnification enables detection of the lensed supernova in $i$-band. The peak magnitude (without magnification) is calculated assuming type Ia supernovae with light curves simulated as outlined in the next section. Given a particular lensing configuration, the in-built lens equation solver in \textsc{lenstronomy} can be used to compute the positions of the multiple lensed images of the supernova and their associated magnifications. We also account for a microlensing \citep{dobler2006microlensing} effect, whereby the strongly lensed supernova is microlensed by the stars in the lens galaxy, by incorporating a stochastic perturbation in the computed magnification. The perturbation does not vary in time and thus, it neglects the evolution of supernova photosphere, which may additionally modify the shape of light curves and change the actual time delays measured from the arrival time of the light curve peaks \citep{pierel2019turning}. However, keeping in mind that the secondary images of lensed supernovae expected in YSE are just around or below the detection limit, the adopted microlensing model is sufficient for the purpose of generating reliable images in the conditions limited by the survey's design. The stochastic perturbation in the macrolens magnification field is computed assuming a Gaussian distribution with a standard deviation of 0.05, similar to what was assumed in \citet{park2021largescale} for generating simulated images of gravitationally lensed AGN.

In this work, we consider only $i$-band images, and we therefore adopt the zero-point magnitude for $i$-band \citep{tonry2012panstarrs1} in line with the Pan-STARRS1 specifications. Unlensed supernovae have a magnification of unity and are rendered at the position of the host galaxy. Fig.~\ref{fig:ensemble_images} depicts a random set of simulated realisations of lensed and unlensed supernovae, generated in accordance with YSE $i$-band specifications so as to closely resemble the actual YSE observations. An analogous procedure is used to simulate LSST $i$-band observations with the image properties specified in Table~\ref{tab:YSE_specs}.

\subsection{Simulating time-series images}
\label{time_series_sim_section}

In order to closely emulate the YSE data output, which consists of multiple images taken at different observational epochs, we generate time series of images. Compared to single-epoch observations, this provides the advantage of fully capturing the multiplicity of the lensed sources, even when the multiple lensed images do not all appear simultaneously.

We model the supernova variability using synthetic light curves. In this work, we only consider type Ia supernovae, since their characteristic light curves are easy to model and they constitute a large fraction ($\sim 40 \%$) of the predicted Pan-STARRS lensed supernova population \citep{wojtak2019magnified}. The light curves are simulated using \textsc{SNCosmo}\footnote{\url{https://sncosmo.readthedocs.io/en/stable/}} \citep{barbary2016sncosmo} and its built-in parametric light curve model SALT2 \citep{guy2007salt2}, which takes as input an amplitude parameter $x_0$, stretch parameter $x_1$, and a colour parameter $c$. We sample the $x_1$ and $c$ parameters from asymmetric Gaussian distributions \citep{scolnic2016measuring} that have been derived for the Pan-STARRS1 data release \citep{rest2014cosmological}. Then, the Tripp formula \citep{tripp1998two} provides a relation for the absolute $B$-band peak magnitude $M_{\textrm{B}}$ that a type Ia supernova, based on its stretch and colour parameters, is expected to have:
\begin{equation}
M_{\textrm{B}} = - \alpha  x_1 + \beta  c + M_{\textrm{abs}},
\end{equation}
where $M_{\textrm{abs}} \sim \mathcal{N}(-19.43, 0.12)$ is the expected absolute magnitude of a supernova with $x_1 = c = 0$, following the standard Planck 2018 $\Lambda$CDM calibration \citep{planck2018cosmo}. The coefficients $\alpha = 0.14$ and $\beta = 3.1$ \citep{scolnic2016measuring} specify the correlation of absolute magnitude with the stretch and colour parameters, respectively. We assume a Milky Way dust extinction model \citep{fitzpatrick1999correcting}, with optical total-to-selective extinction ratio $R_{\textrm{V}} = 3.1$ and low $E(B-V)$ values, since YSE chooses fields with high Galactic latitude and low Milky Way extinction. The adopted distribution for Milky Way dust extinction and other input parameters to the light curve generation routine are provided in Table~\ref{tab:param_distributions}.

After the light curves have been generated, they are used to model the apparent magnitude (including $K$-corrections) for both the lensed and unlensed sources. For the lensed systems, the brightness of each image follows the variability of the light curve, with a correction for the computed magnification, stochastic microlensing perturbations and time delays. Finally, we transform the apparent magnitude to data counts per second using the zero-point magnitude of the instrument, which, when multiplied with the exposure time, yields the amplitude in the desired units for \textsc{lenstronomy}. The observational epochs for each configuration are sampled with a 6-day cadence in the time interval that the observed supernova is brighter than the limiting Pan-STARRS magnitude, resulting in a varying number of images per system. The temporal evolution of a given lensed source, as observed under YSE-like settings, is indicated in Fig.~\ref{fig:temporal_evolution_images}. The Mock Lenses in Time software package \citep{vernardos2022simulating}, published towards the end of our study, presents another alternative for generating time-series images and incorporates a more rigorous treatment of microlensing.

To optimise the three neural networks, we generate a balanced training data set, i.e. with around 50 per cent lensed and unlensed sources, containing a total of 40000 realisations, where 20 per cent is kept for validation. For model performance evaluation, we generate a separate test set consisting of two subsets containing purely lensed and unlensed sources, respectively, with each subset holding 10000 realisations. The images are simulated as detailed in the previous section, along with an extra layer of complexity to incorporate the temporal evolution aspect in the pipeline. The number of time-series observations for each source will depend on its light curve variation and YSE cadence. A new value of PSF is drawn for each multiple-epoch observation as we do not expect correlations in PSF due to the low YSE cadence. Our simulated data set consists solely of sources whose apparent magnitudes are within the YSE ($i$-band) limiting magnitude. To train the single-epoch model, single-epoch observations are randomly drawn from the time-series observations of each source. For the compressed temporal model, we employ the smooth manifold extraction technique, as described in the following section, to combine the series of an arbitrary number of images per source into a single informative image. The spatio-temporal model, in contrast, is trained using all the time-series images of a given source, with arbitrary time intervals between the single-epoch observations.

For real data applications, the preprocessing of the images, as obtained from the YSE data pipeline, before feeding them to our deep learning machinery, is straightforward as the YSE pipeline already provides difference images that are centred on the transient. The only requirement is to extract the central portion extending over 12$^{\prime \prime}$, in accordance with the image stamp size of the images in our training set.

\subsection{Smooth manifold extraction}
\label{SME_section}

\begin{figure*}
	\centering
		{\includegraphics[width=\hsize,clip=true]{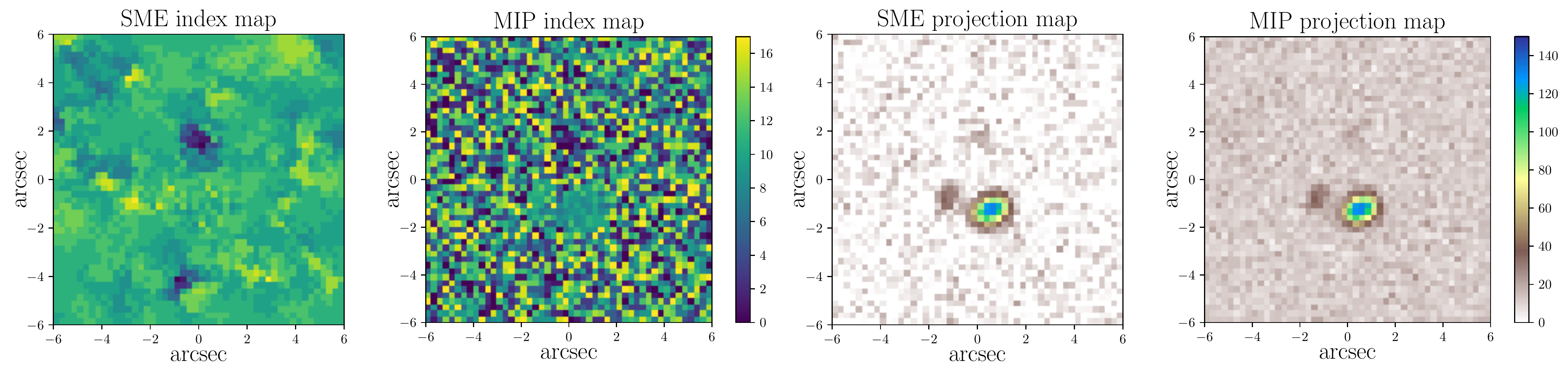}}
	\caption{Comparison between smooth manifold extraction (SME) and maximum intensity projection (MIP) in terms of their respective temporal index maps and corresponding projection maps. The SME index map, unlike the highly discontinuous MIP version, smoothly extracts the signal embedded in the stack of time-series images, thereby enforcing spatial consistency. Consequently, the compressed SME image representation shows a significantly improved contrast with respect to its MIP counterpart. Moreover, the background properties of the SME image are not influenced by the number of time-series images for a given transient, such that this does not artificially bias the neural classifier during training.}
	\label{fig:SME_MIP_comparison}
\end{figure*}

The smooth manifold extraction \citep[SME,][]{shihavuddin2017SME} algorithm is a technique that allows the extraction of the signal embedded in a stack of 2D images, while preserving the local spatial relationships in the original volume of data. It was originally proposed in the field of medical image processing to compress the series of images obtained via fluorescence microscopy into a single 2D representation. The underlying motivation was to improve upon the standard maximum intensity projection (MIP) method that extracts a discontinuous layer of pixels from a 3D stack of images, thereby resulting in artefacts in the final compressed or projected image that may lead to misleading interpretations.

In this work, we implement a variant of the original SME algorithm, tailored for our specific problem, in order to scan the time-series images of a given source and extract the signal without modifying the image properties, such as the background noise, of single-epoch observations. In contrast, the implementation of MIP is straightforward as it consists in retrieving the level of maximum intensity along the temporal axis for each $(x,y)$ spatial position, with the map of levels referred to as the index map $\mymat{Z}$. MIP is not suited to our classification problem as it biases the network learning process. This is due to the fact that the simulated lensed sources have, on average, a higher number of time-series snapshots than the unlensed ones, such that MIP results in distinct background characteristics for the two cases.

The rationale of the SME optimisation routine is to fit a smooth 2D manifold onto the foreground signal, while disregarding the background, thereby propagating the index map from the foreground to the local background. To this end, we constrain each pixel in the index map by minimising the distance between the highest intensity region and the local variance of the index map. The former ensures foreground proximity of the index map, while the latter enforces its smoothness. The optimal index map is, therefore, obtained by minimising the cost function:
\begin{equation}
    \mymat{Z}_{\mathrm{SME}} = \argmin{\mymat{Z}} \!\!\! \sum_{(x,y)} \!\! \mathcal{W}(x,y) | \mymat{Z}_{\mathrm{max}}(x,y) - \mymat{Z}(x,y) | + \sigma_{\mymat{Z}}(x,y) ,
    \label{eq:SME_cost_function}
\end{equation}
where $\mymat{Z}_{\mathrm{max}}$ corresponds to the MIP index map and $\sigma_{\mymat{Z}}$ is the local spatial standard deviation computed for a $3 \times 3$ window centred on $(x,y)$. The $\mathcal{W}$ operator assigns a weight to each pixel to quantify whether it encodes a signal. As weighting scheme, we opt for the following \texttt{softplus} function:
\begin{equation}
    \mathcal{W}[ f_{i \in (x,y)} ] = \log \big\{ 1 + \sqrt{a \exp{[k(f_i - b)]}} \big\}/k ,
    \label{eq:SME_weighting_operator}
\end{equation}
where the constants are set to $a = 10^{-3}$, $b = 30$ and $k = 0.125$, on the basis of numerical experiments. Finally, we adopt a tolerance threshold $\epsilon$ that is sufficiently stringent to allow the cost function to converge. The numerical implementation of our variant of the SME algorithm, as outlined in Algorithm~\ref{alg:SME}, employs a distinct weighting strategy and convergence scheme with respect to the original implementation \citep{shihavuddin2017SME}, and results in smooth convergence of the cost function. A comparison of the temporal index maps for SME and MIP, and their respective compressed image representations, is illustrated in Fig.~\ref{fig:SME_MIP_comparison}.

\begin{algorithm}[t]
\begin{algorithmic}[1]
  \Procedure{\textsc{SME}}{$\bm{\mathcal{I}}$} \\
  \Comment{Input is an image stack $\bm{\mathcal{I}}$ with dimensions $W \times H \times D$}
  \State $\mymat{Z}_{\mathrm{max}}(x,y) = \argmax{z} \bm{\mathcal{I}}(x,y,z) $ \Comment{MIP index map}
  \State $\mymat{I}_{\mathrm{MIP}} = \bm{\mathcal{I}}[x,y,\mymat{Z}_{\mathrm{max}}(x,y)]$ \Comment{MIP 2D image}
  \State $\mymat{Z}_0 = \mymat{Z}_{\mathrm{max}}(x,y)$ \Comment{Initialise $\mymat{Z}$ with MIP index map}\\
  \Comment{Initial step size ($T_0$), step factor ($\Delta T$), tolerance ($\epsilon$)}
  \State $i = 1$, $T_0 = D/100$, $\Delta T = 0.99$, $\epsilon = 5\times10^{-3}$
  \While{$\left\| \mymat{Z}_{i} - \mymat{Z}_{i-1} \right\| / \left\| \mymat{Z}_{i} \right\| > \epsilon$}
    \For{$\forall (x,y) \in W \times H$} \\
        \Comment{All quantities below are a function of $(x,y)$}
        \State $\Delta \mymat{Z}_i \in \{ -T_i, 0, T_i \}$
        \State $\Ztilde \gets \mymat{Z}_{i-1} + \Delta \mymat{Z}_i$
        \State $\mymat{Z}_i \gets \argmin{\Ztildee} \mathcal{W}(\mymat{I}_{\mathrm{MIP}}) | \mymat{Z}_{\mathrm{max}} - \Ztilde | + \sigma_{\Ztildee}$
    \EndFor
    \State $T_i \gets T_i \times \Delta T$
    \State $i \gets i + 1$
  \EndWhile
    \State \Return{$\bm{\mathcal{I}}[x,y,\mymat{Z}_i(x,y)] \equiv \mymat{I}_{\mathrm{SME}}$} \Comment{SME 2D image}
  \EndProcedure
\end{algorithmic}
\caption{\label{alg:SME} Smooth manifold extraction}
\end{algorithm}

\subsection{Conventional neural networks}
\label{conventional_NN_section}

Before outlining the basic principles underlying Bayesian neural networks, as relevant to our study, a concise description of conventional neural networks is first in order, so as to highlight the contrast with the former.

In a supervised classification problem, the neural network, $\mathbb{NN}(\myvec{\omega}, \myvec{\gamma}): \myvec{\mathcal{D}} \rightarrow \myvec{\tau}$, as an arbitrarily complex and flexible model, maps some input data $\myvec{\mathcal{D}}$ to a prediction of the desired label $\myvec{\tau}$ associated with the data, where $\myvec{\omega}$ and $\myvec{\gamma}$ correspond to the trainable model parameters, known as {\it weights}, and hyperparameters (e.g. network architecture, weights initialisation, type of activation and loss functions), respectively. The weights are optimised during training to minimise a certain {\it loss function}. The latter is a differentiable function of the data and the weights, quantifying how well the network's output matches the associated label of the input data, and is often taken to be the negative log-likelihood. The training routine is, therefore, equivalent to determining a maximum likelihood estimate of the weights, resulting purely in single point predictions by the neural network.

The network's prediction accuracy and overall efficacy is generally evaluated via a performance metric, with the conventional choice for a classification task being the {\it accuracy}, i.e. the fraction of correctly assigned labels. Nevertheless, this is not sufficient to assess the reliability of any individual network prediction and it is, hence, imperative to quantify the confidence associated with a given prediction. For a standard neural classifier, the final output layer is an $N$-dimensional vector, corresponding to a classification problem with $N$ classes, whose components sum to unity, such that they can be interpreted as probabilities. The trained network, for a given test realisation, then assigns the label to the one with largest output probability score.

A caveat is that the network probability score should not be interpreted as the confidence in the prediction as it is explicitly dependent on the maximum likelihood estimates of the weights, the training data and the set of hyperparameters adopted during the network training procedure. To quantify the confidence associated with a predicted label, it is essential to account for two classes of uncertainties \citep{kendall2017uncertainties}, namely {\it aleatoric} and {\it epistemic} uncertainties, encompassing the uncertainties in the data and model, respectively. The former describes the inherent uncertainty by virtue of the random nature of the observations of interest, while the latter refers to the limitations of the model to accurately describe the set of observations.

\begin{figure*}
	\centering
		{\includegraphics[width=\hsize,clip=true]{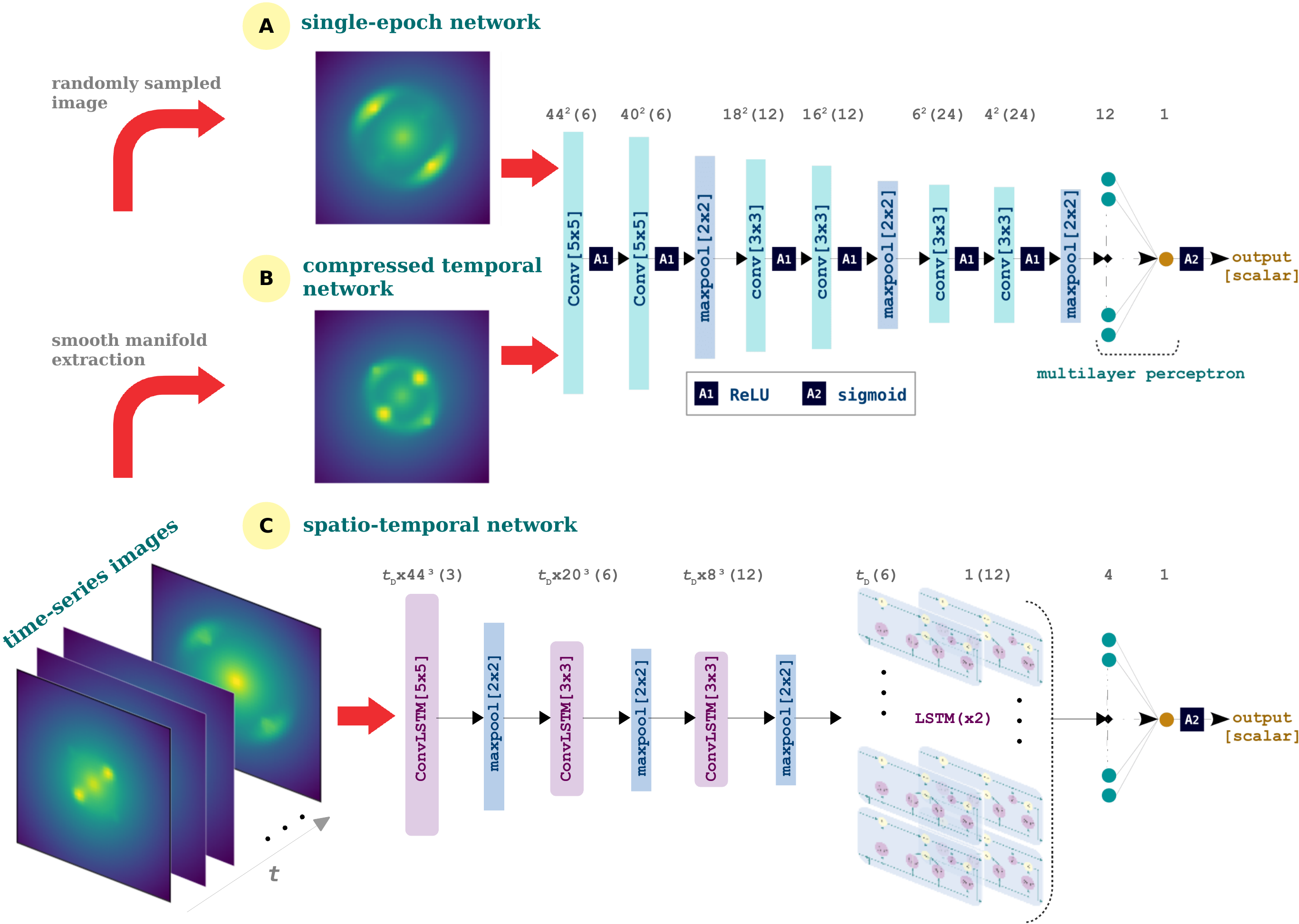}}
	\caption{Schematic representation of our Bayesian neural classifiers. The three models have distinct inputs, but with a similar scalar output, corresponding to a probability score, that is used for classification. The spatio-temporal model, unlike the single-epoch and compressed temporal models, optimally exploits the information characterised by the temporal evolution of the astrophysical transient. We combine our neural classifiers with variational inference to provide a confidence score, thereby quantifying approximate Bayesian uncertainties associated with each network prediction. The dimensions of the image slices, resulting from the various convolutional, maxpooling or LSTM operations, are indicated above the architecture schematics, with the number of feature maps per layer displayed in parentheses and $t_{\mathrm{D}}$ denoting the temporal dimension (number of time-series images per source) for the spatio-temporal network. Dropout masks are applied after each convolutional or dense layer, except for the final output layer. The AI engines have a relatively low model complexity with $\mathcal{O}(10^4)$ trainable parameters.}
	\label{fig:classifier_schematic}
\end{figure*}

\subsection{Bayesian neural classifiers}
\label{bayesian_NN_section}

To overcome the limitation of point weight estimates and quantify the uncertainty associated with the model weights, neural networks combined with variational inference, commonly designated as {\it Bayesian neural networks} \citep{mackay1992practical, neal2012bayesian, charnock2020bayesian}, cast the model parameters as probability distributions and subsequently marginalise the network's output over these distributions within a Bayesian statistical framework to yield a network score posterior, thereby quantifying the uncertainty inherent in model selection. Hence, a trained Bayesian neural network represents an ensemble of networks, which allows the uncertainty on a specific classification to be quantified. While there is a panoply of distinct approaches and implementations, recent studies demonstrate the potential of such techniques in providing reliable classification uncertainties \citep{mancarella2022seeking, killestein2021transient}.

The network probability score captures, to some extent, a measure of the aleatoric uncertainties due to the noise intrinsic to the input data set. To more adequately account for this source of uncertainties, we must ensure that the training data set is representative of the distribution of possible observations, as enforced by sampling all the input parameters relevant for the image generation pipeline from their respective plausible distributions (cf. Table~\ref{tab:param_distributions}). Next, to account for the epistemic uncertainties associated with the choice of neural network’s weights, the most common approach is to replace each network weight by a parametrised distribution to eventually infer the posterior distribution of the weights conditional on the input data during training via Bayes identity:
\begin{equation}
    \mathcal{P}(\omega | \mathcal{D}) = \frac{\mathcal{L}(\mathcal{D}|\omega) \mathcal{P}(\omega)}{\mathcal{P}(\mathcal{D})} .
    \label{eq:bayes_identity}
\end{equation}
Since the posterior $\mathcal{P}(\omega | \mathcal{D})$ is intractable in practice, variational inference is typically used to approximate the posterior by a variational distribution $q_{\theta}(\omega)$, where $\theta$ characterises an ensemble of distributions. The training objective is then to ensure that the variational distribution $q_{\theta}(\omega)$ matches the posterior distribution $\mathcal{P}(\omega | \mathcal{D})$ as closely as possible, with the measure of similarity generally quantified by the K\"ullback-Leibler (KL) divergence \citep{kullback1951information}:
\begin{equation}
    \mathrm{KL}[ q_{\theta}(\omega)|| \mathcal{P}(\omega | \mathcal{D})] \equiv \int \mathrm{d} \omega \; q_{\theta}(\omega) \log \left[ \frac{q_{\theta}(\omega)}{\mathcal{P}(\omega | \mathcal{D})} \right] .
    \label{eq:KL_divergence}
\end{equation}
Using Bayes identity, the above loss function can be expressed in terms of the likelihood $\mathcal{L}(\mathcal{D}|\omega)$ and the prior distribution $\mathcal{P}(\omega)$ for the network weights \citep{blundell2015weight}:
\begin{align*}
    \mathrm{KL}[q_{\theta}(\omega)||\mathcal{P}(\omega | \mathcal{D})]  = &\mathrm{KL}[q_{\theta}(\omega)||\mathcal{P}(\omega)] \\ &- \mathbb{E}_{q_{\theta}(\omega)} [\log \mathcal{L}(\mathcal{D}|\omega) ] + K , \numberthis
    \label{eq:VI_loss}
\end{align*}
where $\mathbb{E}_{q_{\theta}(\omega)}$ denotes the expectation under the variational distribution, with the constant $K$ resulting from the Bayesian evidence and the second term being the standard negative log-likelihood. We can obtain a Monte Carlo estimate of this loss by sampling the weights from the variational distribution, $\omega \sim q_{\theta} (\omega | \mathcal{D})$. In simpler terms, variational inference implies the assumption of the form of the posterior distribution of the weights and the use of an optimisation routine to find the assumed probability distribution that is closest to the true posterior. This assumption simplifies the computation, resulting in some level of tractability.

In this work, we adopt the variational inference technique known as Monte Carlo (MC) Dropout \citep{gal2015bayesian}. This method derives from the dropout regularisation approach \citep{srivastava2014dropout}, whereby a fraction of model weights are randomly set to zero during each training step, effectively lowering the number of model parameters to ultimately mitigate risks of overfitting. Dropout is traditionally deactivated at inference time, such that network predictions are deterministic. It was subsequently shown that training and evaluating neural networks with dropout is equivalent to implementing approximate Bayesian inference \citep{gal2015dropout}, whereby multiple model evaluations can be used to perform Monte Carlo integration of the posterior distribution. Formally, MC Dropout implies that a Bernoulli distribution is assumed as the variational distribution. Under this formulation, it can be shown that including an $\ell_2$ regularisation term, i.e. having Gaussian priors over the network weights, approximately entails computing the KL divergence with respect to an implicit prior \citep{gal2015bayesian}. Consequently, training a neural network with dropout masks and $\ell_2$ weight regularisation minimises the loss from equation~\eqref{eq:VI_loss}, such that it is feasible to perform proper variational inference without intensifying the computational workload \citep{charnock2020bayesian}.

We use the posterior predictions obtained via the MC Dropout method to derive a confidence score, as a single summary statistic, that encapsulates the epistemic uncertainty for each network classification. To this end, we make use of the information entropy $\mathcal{H}$ in the binary classification context \citep{houslby2011bayesian}:
\begin{equation}
    \mathcal{H}(p) = -p \log_2 p - (1 - p) \log_2 (1 - p) ,
    \label{eq:information_entropy_binary}
\end{equation}
where $p$ is the network output probability. The neural classifier confidence $\mathcal{C}$ can then be computed from the average entropy of $N_{\mathrm{MC}}$ posterior samples, as follows \citep{killestein2021transient}:
\begin{equation}
    \mathcal{C} = 1 - \frac{1}{N_{\mathrm{MC}}} \sum^{N_{\mathrm{MC}}}_{i=1} \mathcal{H}_i ,
    \label{eq:classifier_confidence_binary}
\end{equation}
where $\mathcal{H}_i$ denotes the binary entropy of the $i$\textsuperscript{th} sample, yielding a confidence score in the range $[0,1]$. While this approach provides uncertainties that are correlated with the network probability scores, the dispersion resulting from multiple samples is still informative, with ten MC samples found to improve classification performance relative to a deterministic network \citep[cf. Fig.~6 in][]{killestein2021transient}. In our work, we set $N_{\mathrm{MC}} = 50$.

Our choice of variational method is driven primarily by the intended practical utility of our AI tool for sifting through huge volumes of plausible lensed transients. Having a separate confidence score for a specific source, in addition to the network probability score, is informative for both human vetting and automated filtering pipelines. The numerical implementation is also straightforward, with the extra requirements being the inclusion of dropout masks after the convolutional and dense layers, except for the final output layer, with $\ell_2$ weight regularisation. The network training routine is also unchanged with respect to a conventional neural network, with the additional computational workload being the multiple model evaluations at inference time for a given input image. The MC Dropout method is robust as long as we restrict our neural network to low model complexity and employ a large training data set, such that it is justified to assume minimal dropout rates and $\ell_2$ regularisations. A more rigorous way of quantifying epistemic uncertainties is by training an ensemble of networks with randomly drawn architectures, variational distributions and hyperparameters, but this is beyond the scope of the practical tool developed in this work.

\subsection{Convolutional neural network}
\label{CNN_section}

The single-epoch and compressed temporal models are cast as a convolutional neural network \citep[CNN,][]{lecun1995convolutional, lecun2015deep}, composed of convolutional and max-pooling layers, and eventually a fully-connected layer leading to the final output layer. Each convolutional layer has a set of kernels that learn to detect a certain type of feature present in the network input. We made use of 2D kernels whose weights are first randomly initialised and subsequently updated during the network training routine. Each convolutional kernel yields a \textit{feature map} that indicates the strength and location of the feature detected by the kernel. The fully-connected layers consist of units known as {\it neurons}. Each neuron in a given layer is connected to every neuron in adjacent layers and applies a non-linear {\it activation function} to a weighted sum of its inputs to return a single output.

For our classification problem, the input data set for the single-epoch and compressed temporal models comprises a collection of lensed and unlensed supernova images (single-epoch or compressed time-series representation, respectively), while the ground truth labels are binary in nature. As depicted in Fig.~\ref{fig:classifier_schematic}, the CNN architecture consists of three blocks of two convolutional layers employing 6, 12 and 24 kernels of sizes $5 \times 5$, $3 \times 3$ and $1 \times 1$, respectively, with unit stride and zero-padding, followed by a max-pooling operation. The first layer learns features on scales of $1.25^{\prime \prime}$. With the successive stacking of convolutional layers, low-level local features gleaned in the initial layers are merged into high-level global features by the following layers, rendering the network sensitive to features on larger scales. This allows both local and global information to propagate through the deep learning machinery. We applied the non-linear rectified linear unit (\texttt{ReLU}) \citep{nair2010ReLU} activation function, $f(x) = \mathrm{max}(0,x)$, to every feature map, except for the output layer. The feature maps are then fed to a max-pooling layer that extracts the maximum value over $2\times2$ non-overlapping regions of the feature maps for dimensionality reduction. The output from the final convolutional layer and its corresponding pooling layer is flattened into a 1D vector and fed to two fully-connected layers with 12 and 4 neurons, respectively. The final output layer has a \texttt{sigmoid} activation to yield outputs in the range $[0,1]$, as per the convention for networks designed for binary classification.

\subsection{Convolutional recurrent neural network}
\label{ConvLSTM_section}

Recurrent neural networks \citep[RNNs,][]{lipton2015RNNs} consist of neurons, where the output of each neuron in a given layer is not only connected to the next layer but is also fed as input to itself. The temporal behaviour of such an architecture renders RNNs suitable for learning sequential models or time series. Long short-term memory \citep[LSTM,][]{hochreiter1997LSTM} networks are a specific type of RNNs, which build upon these models by encoding a new element in the neurons, known as the {\it memory cell} ($c_t$). The role of the latter is to accumulate the information from the previous inputs $x_{t-1}$, thereby allowing the neuron to recall the information that it has already processed. By relying on a gating mechanism, the neuron decides whether to accumulate the new input $x_t$ in the memory cell $c_t$ and whether it should forget the prior state $c_{t-1}$ and propagate $c_t$ to the {\it hidden state} $h_t$. LSTM is implemented via the following equations:
\begin{align}
\mathrm{Input \ gate} \;\;\;\; & i_t = \sigma (W_i x_t + b_{ii} + U_i h_{t-1} + b_{hi}) \label{eq:input_gate} \\
\mathrm{Forget \ gate} \;\;\;\; & f_t = \sigma (W_f x_t + b_{if} + U_f h_{t-1} + b_{hf}) \label{eq:forget_gate} \\
\mathrm{Cell \ gate} \;\;\;\; & g_t = \tanh (W_g x_t + b_{ig} + U_g h_{t-1} + b_{hg}) \label{eq:cell_gate} \\
\mathrm{Output \ gate} \;\;\;\; & o_t = \sigma (W_o x_t + b_{io} + U_o h_{t-1} + b_{ho}) \label{eq:output_gate} \\
\mathrm{Cell \ state} \;\;\;\; & c_t = f_t \times c_{t-1} + i_t \times g_t \label{eq:cell_state} \\
\mathrm{Hidden \ state} \;\;\;\; & h_t = o_t \times \tanh(c_t) ,
\label{eq:hidden_state}
\end{align}
where $x_t$ is the input at time $t$, $\sigma$ is the recurrent (\texttt{sigmoid}) activation function, $W$ and $U$ are the trainable weights, while $b$ indicates the bias terms and $\times$ denotes element-wise product. $h_{t - 1}$ is the hidden state at the prior time step or its initialised state at $t=0$, and similarly for $c_{t - 1}$. In essence, $h_t$ and $c_t$ encapsulate the short (fast) and long (slow) correlations in the data, respectively, and hence, correspond to the cell's slow and fast hidden states. This dichotomy alleviates the issue of vanishing or exploding gradients that plague vanilla RNNs.

\begin{figure}
	\centering
		{\includegraphics[width=\hsize,clip=true]{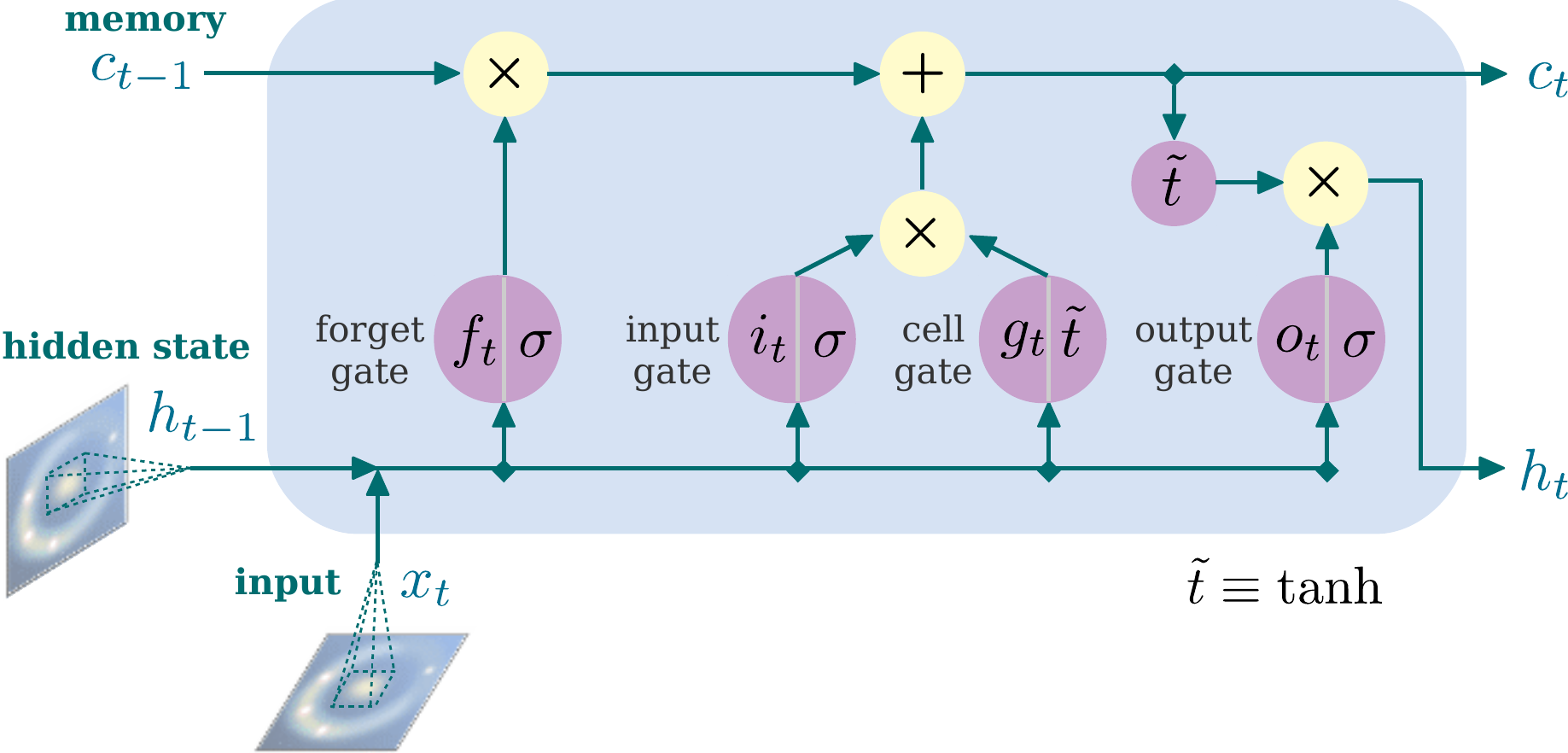}}
	\caption{Schematic representation of the ConvLSTM cell, illustrating the mechanism for updating the cell's memory ($c_t$) and hidden ($h_t$) states for a single time step.}
	\label{fig:ConvLSTM_cell}
\end{figure}

Convolutional LSTM \citep[hereafter ConvLSTM,][]{shi2015ConvLSTM} is an extension of the LSTM network tailored to exploit spatially-correlated features, in addition to temporal correlations, in data. The main change is that the 1D vectors representing the inputs, cell states and gates in equations~\eqref{eq:input_gate}--\eqref{eq:hidden_state} are now 3D tensors, and the products with the weight matrices $W$ and $U$ now represent convolutions over spatial dimensions, with the updates of the cell's states proceeding as before. A schematic representation of the ConvLSTM cell is displayed in Fig.~\ref{fig:ConvLSTM_cell}.

The ConvLSTM architecture of the spatio-temporal network is illustrated in Fig.~\ref{fig:classifier_schematic}. The input data set contains a collection of time-series images of lensed and unlensed supernovae, with the ground truth labels and final network output being identical to those of the CNN models. The input time-series images are first processed through three ConvLSTM layers, encoding 6, 12 and 24 cells, respectively, combining 2D convolutions in the spatial domain with recurrent LSTMs across the temporal dimension. For dimensionality reduction, max-pooling operations are applied after each ConvLSTM layer, followed by a stack of two LSTM layers, with 6 and 12 cells, respectively, after a flattening step. The final part of the architecture comprises of a fully-connected layer with 4 neurons that leads to the network output.

Some common applications of RNNs involve speech recognition, language translation and video tagging. In the field of astrophysics, the ConvLSTM model was recently used to infer galaxy properties from 21 cm lightcone images \citep{prelogovic2022machine}, with this study also showing that the ConvLSTM network outperforms a standard 3D CNN, where the third dimension incorporates the temporal evolution, for this particular problem.

\subsection{Network implementation and training}
\label{NN_training_section}

We implement our Bayesian neural classifiers using the \textsc{Keras} framework \citep{chollet2015keras} via a \textsc{TensorFlow} backend \citep{abadi2016tensorflow}. To train the neural networks, we use the \textsc{adam} \citep{kingma2014adam} optimiser for robust and reliable convergence, and the binary cross-entropy loss function designed for binary classification problems. We set the learning rate to $\eta=10^{-4}$, along with the default values, $\beta_1=0.9$ and $\beta_2=0.999$, for the first and second moment exponential decay rates, respectively. We use a fixed batch size of 100 samples and train our neural classifiers for around 50 epochs, with the training routine for the CNN and ConvLSTM running to completion in around ten and forty minutes, respectively, on an NVIDIA V100 Tensor Core GPU.

To limit network overfitting, we adopt the early stopping regularisation technique \citep{goodfellow2016deep}, and implement an early stopping criterion of five epochs. To this end, $20\%$ of the training data set is randomly selected to comprise a validation set. Training is terminated when the validation loss ceases to improve for five consecutive epochs, with the weights of the previously saved best fit model restored. We use a marginal dropout probability of $10^{-2}$ between the convolutional and fully-connected layers, along with $\ell_2$ weight regularisation penalty of $10^{-4}$, which provide further regularisation. The inclusion of the dropout masks and $\ell_2$ regularisation, however, is primarily for the purpose of variational inference to quantify the confidence score associated with a given network classification. Our CNN architecture, consisting of 12075 trainable parameters, has a relatively low model complexity and requires negligible regularisation given the large volume of training data. Our ConvLSTM architecture has a comparable model complexity with 16749 trainable weights. We leave any (automated) hyperparameter tuning to a future work.

\section{Results}
\label{results}

The network architecture of our three AI models for finding lensed supernova is illustrated in Fig.~\ref{fig:classifier_schematic}. Both the single-epoch and compressed temporal models employ the same CNN architecture, while the spatio-temporal network is constructed using recurrent convolutional layers \citep{shi2015ConvLSTM} that encode LSTM \citep{hochreiter1997LSTM} cells. We implement our neural classifiers within the MC Dropout \citep{gal2015bayesian} framework to approximately quantify the Bayesian uncertainties associated with a particular classification via a confidence score.

\subsection{Evaluating neural classifier performance}
\label{classifier_performance_evaluation}

\begin{figure}
	\centering
		{\includegraphics[scale=0.675,clip=true]{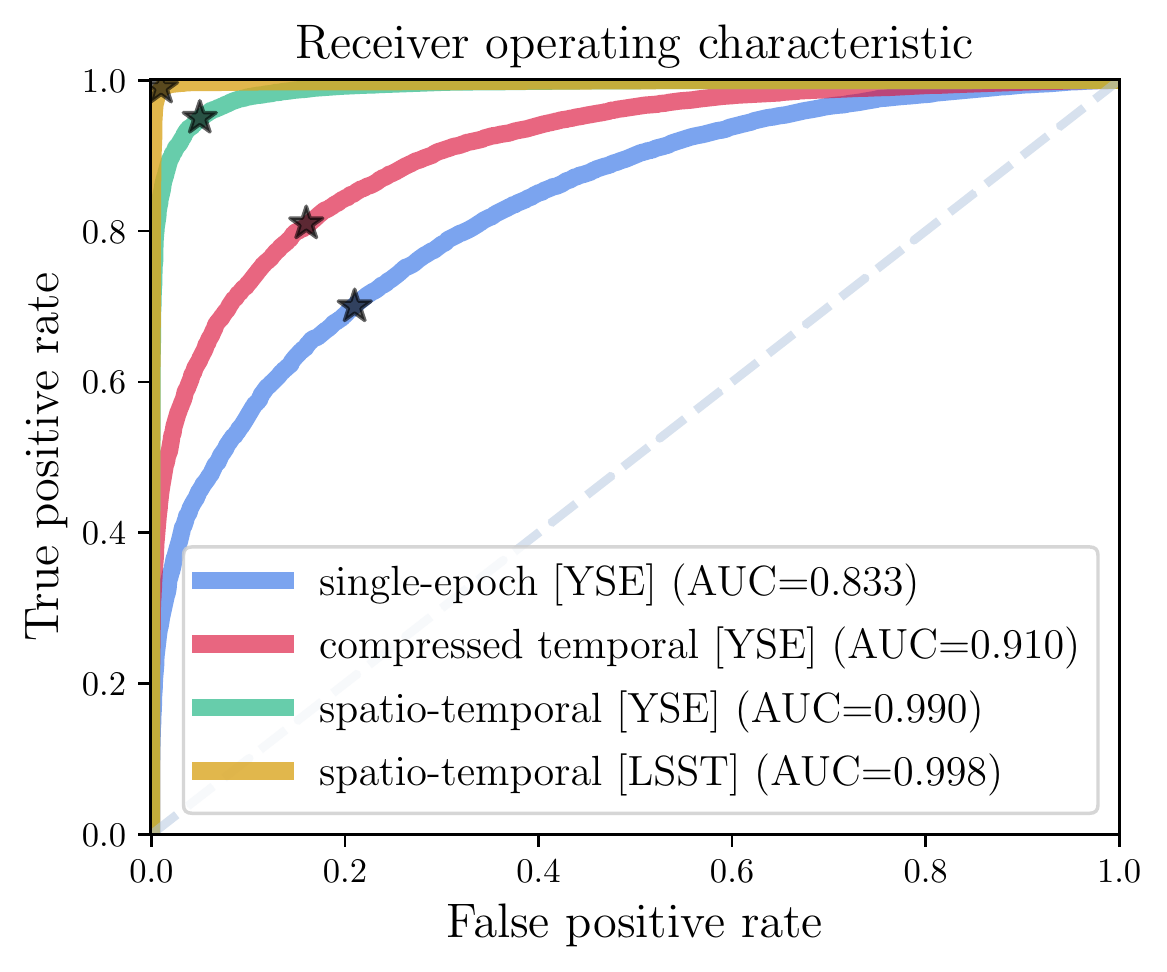}}
	\caption{Receiver operating characteristic curves for the three Bayesian neural classifiers, showing the true positive rate as a function of the false positive rate, both computed for varying detection thresholds in the range [0,1]. The chosen detection threshold of 0.5 used to compute the confusion matrix in Fig.~\ref{fig:confusion_matrix} is indicated using black stars. All classifiers achieve decent AUC scores, with the AUC score of the spatio-temporal model highlighting the progressive improvement in classification efficacy with the addition of temporal information. Its performance on the LSST mock observations is also extremely promising.\vspace{-0.0cm}}
	\label{fig:ROC_curve}
\end{figure}

In order to evaluate the performance of our trained Bayesian neural classifiers on the test set, we employ several standard classification metrics. To quantify the accuracy of our model, we first compute the true positive rate (TPR) and false positive rate (FPR). The former, also known as {\it completeness}, is defined as the fraction of detected lensed sources, whilst the latter, also referred to as the {\it contamination} rate, is defined as the fraction of normal supernovae incorrectly classified as lensed sources:
\begin{equation}
    \mathrm{TPR} = \frac{N_{\mathrm{TP}}}{N_{\mathrm{TP}} + N_{\mathrm{FN}}} , \; \; \;
    \mathrm{FPR} = \frac{N_{\mathrm{FP}}}{N_{\mathrm{FP}} + N_{\mathrm{TN}}} ,
    \label{eq:TPR_FPR}
\end{equation}
where $N_{\mathrm{TP}}$, $N_{\mathrm{FN}}$, $N_{\mathrm{FP}}$ and $N_{\mathrm{TN}}$ denote the number of true positives, false negatives, false positives and true negatives, respectively. These summary statistics are a function of the detection threshold applied to the neural classifier's probability score. The overall performance of the trained classifier is, hence, conventionally assessed by the receiver operating characteristic (ROC) curve. This diagnostic curve depicts the TPR as a function of the FPR and is generated by gradually increasing the detection threshold in the range $[0,1]$, as illustrated in Fig.~\ref{fig:ROC_curve} for our three models. Random predictions by the classifier will produce a diagonal line, with the area under the curve, denoted by the AUC score, equal to 0.5, whilst a perfect classifier will have an AUC score of unity. All three models achieve decent AUC scores, with that of the spatio-temporal one approaching unity, showcasing the gradual improvement of the use of the full time-series images over single-epoch images, as a result of the additional temporal information.

\begin{figure*}
	\centering
    \subfloat[Single-epoch (YSE)]{\includegraphics[width=0.245\hsize]{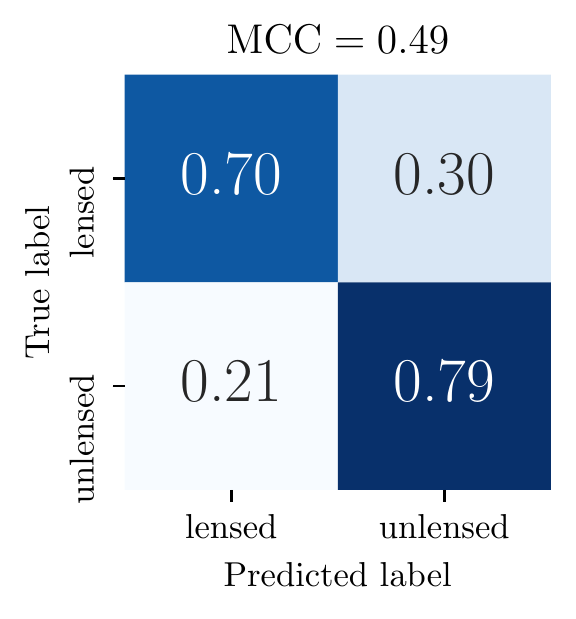}}
    \! \! 
    \subfloat[Compressed temporal (YSE)]{\includegraphics[width=0.245\hsize]{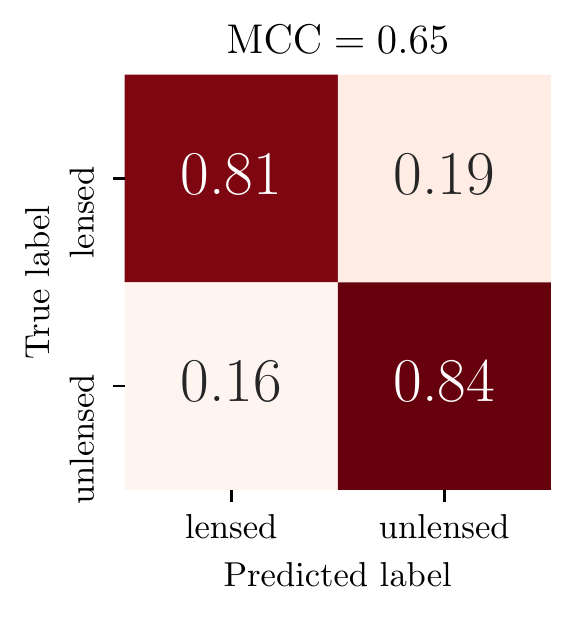}}
    \! \!
    \subfloat[Spatio-temporal (YSE)]{\includegraphics[width=0.245\hsize]{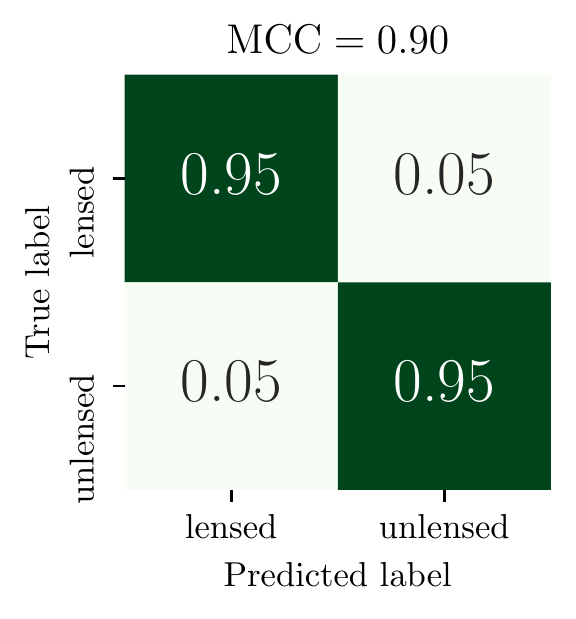}}
    \! \!
    \subfloat[Spatio-temporal (LSST)]{\includegraphics[width=0.245\hsize]{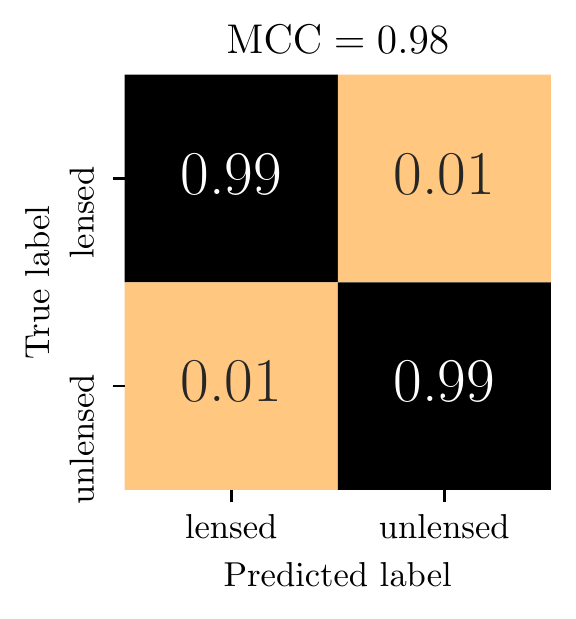}}
	\caption{Confusion matrix showing the classification efficacy of the three models employing single-epoch images, the compressed temporal representation and all the time-series images, respectively, for the YSE set-up. The spatio-temporal model yields an improvement of around 20\% in overall classification accuracy relative to the single-epoch model, with 99\% accuracy for LSST images.}
	\label{fig:confusion_matrix}
\end{figure*}

In this work, we adopt the default value of 0.5 for the detection threshold for both of our models. Given this threshold, we can visualise the overall classification accuracy using the confusion matrix, as shown in Fig.~\ref{fig:confusion_matrix}. The confusion matrix describes the percentage of samples from each class that are accurately classified and simultaneously expresses that of erroneous classifications. As a robust measure of the quality of binary classifications, the Matthews correlation coefficient (MCC) is a commonly used metric to summarise the confusion matrix:
\begin{equation}
    \mathrm{MCC} = \frac{N_{\mathrm{TP}} N_{\mathrm{TN}} - N_{\mathrm{FP}} N_{\mathrm{FN}}}{\sqrt{(N_{\mathrm{TP}} + N_{\mathrm{FP}}) (N_{\mathrm{TP}} + N_{\mathrm{FN}}) (N_{\mathrm{TN}} + N_{\mathrm{FP}}) (N_{\mathrm{TN}} + N_{\mathrm{FN}})}} ,
    \label{eq:MCC}
\end{equation}
with the limiting values of $\mathrm{MCC} = \{-1,1\}$ corresponding to predictions in total disagreement and perfect agreement, respectively, with observations, while $\mathrm{MCC} = 0$ implies random classifier predictions. The MCC scores for our models are given in Fig.~\ref{fig:confusion_matrix}, further demonstrating the improved classification efficacy with the inclusion of temporal information.

\begin{figure}
	\centering
		{\includegraphics[scale=0.64,clip=true]{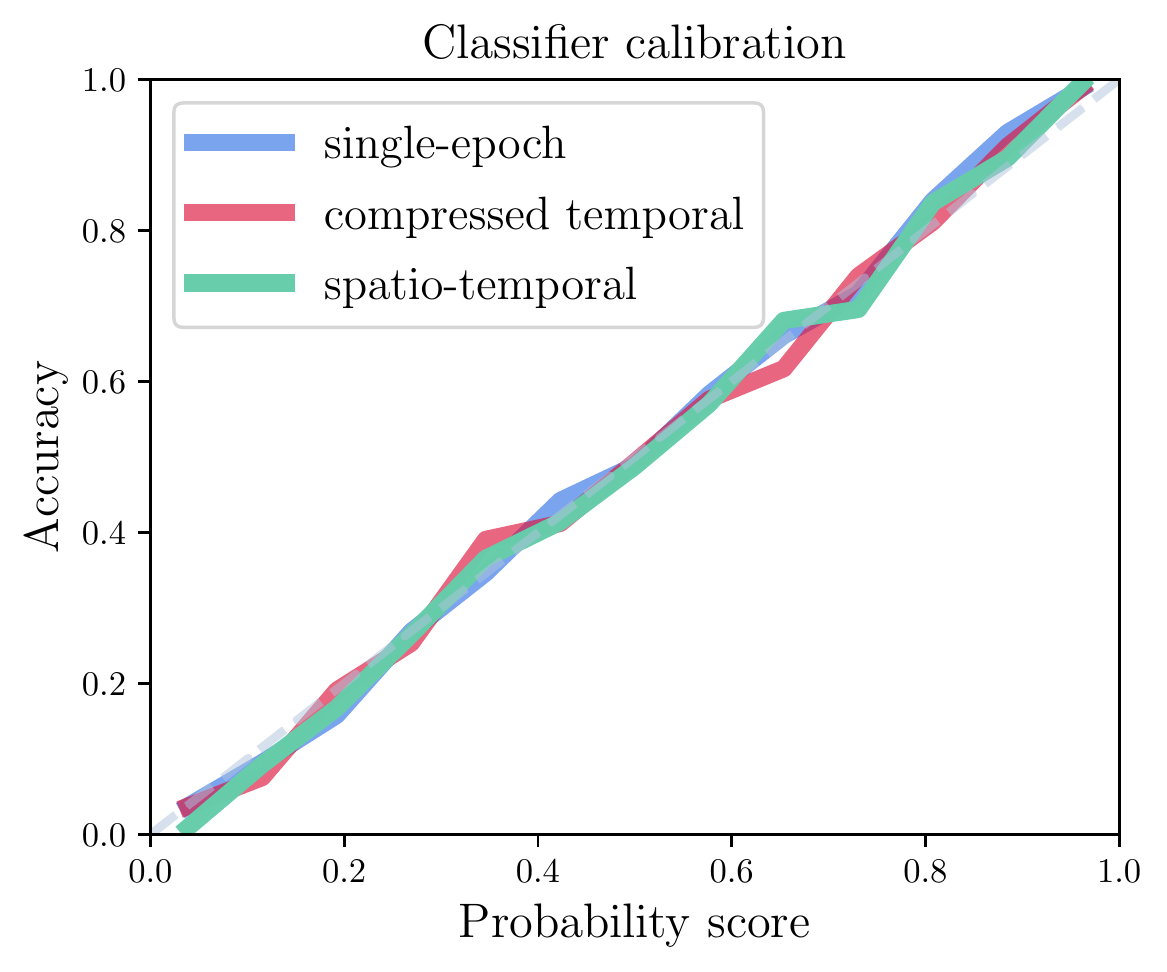}}
	\caption{Classifier calibration indicating how well the network probability score corresponds to probability. The diagonal dashed line implies perfect calibration, i.e. a perfect match between the probability score and accuracy. The calibration curves for the three Bayesian neural classifiers show a remarkable degree of calibration, such that it is justified to use the network probability score as a proxy for the probability that a source is lensed.}
	\label{fig:classifier_calibration}
\end{figure}

\begin{figure}
	\centering
		{\includegraphics[width=\hsize,clip=true]{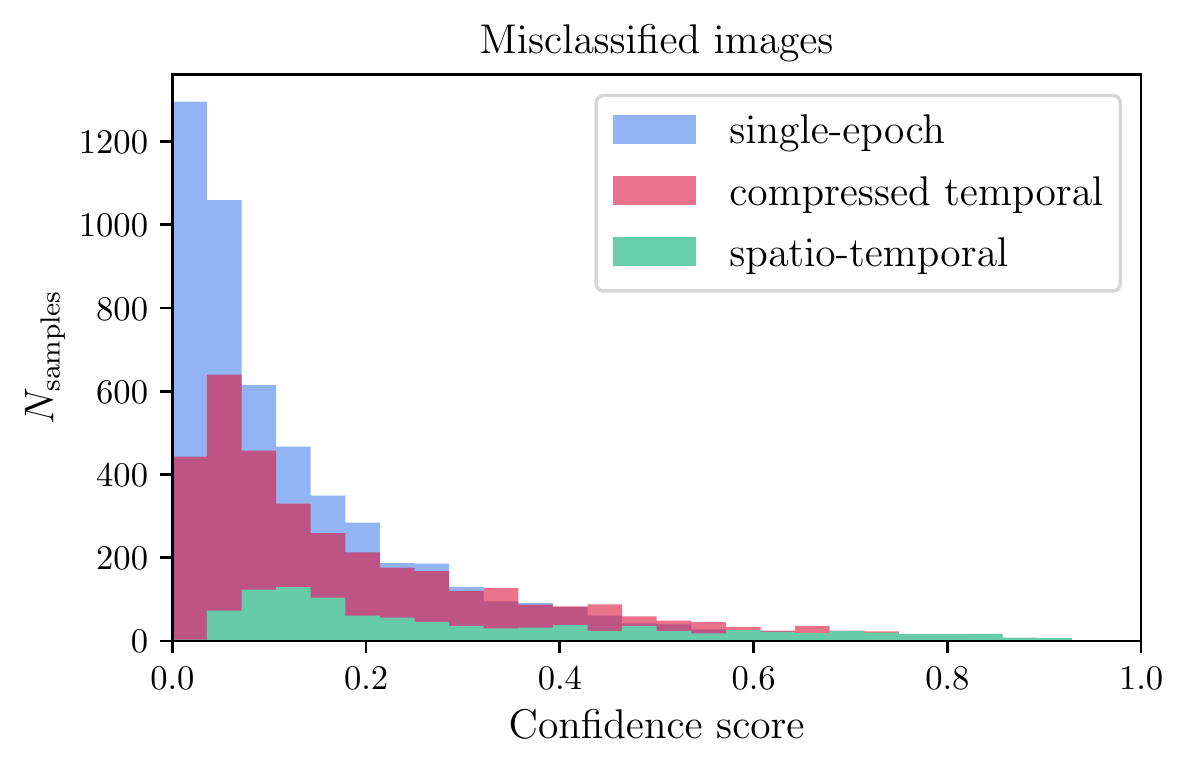}}
	\caption{Distribution of confidence scores for misclassified (false positives and false negatives) images. On average, for all three models, only around 10\% of wrongly classified sources have a confidence score larger than 0.5, thereby demonstrating the reliability of this metric to reflect the neural network uncertainty associated with a given classification.}
	\label{fig:confidence_misclassified_images}
\end{figure}

An essential aspect to consider when using a neural classifier is whether its output, i.e. the network probability score, accurately reflects the probability of a source being lensed or unlensed. In order to verify how closely the network's output correlates with probability, we perform a classifier calibration test in Fig.~\ref{fig:classifier_calibration}. All three models display nearly perfect calibration, implying that the network probability scores may be interpreted as the probability that a certain source is lensed.  Moreover, our Bayesian neural classifiers quantify the reliability of the network classifications via a confidence score, computed from the average information entropy of an ensemble of Monte Carlo posterior samples from the trained networks (cf. equation~\eqref{eq:classifier_confidence_binary} in Section~\ref{methods}). The distribution of such scores associated with wrongly classified transients depicted in Fig.~\ref{fig:confidence_misclassified_images}. We find that there is a clear trend of erroneously classified sources having very low confidence scores, thereby justifying the utility of this additional metric when evaluating potential lensed supernovae.

\begin{figure}
	\centering
		{\includegraphics[scale=0.60,clip=true]{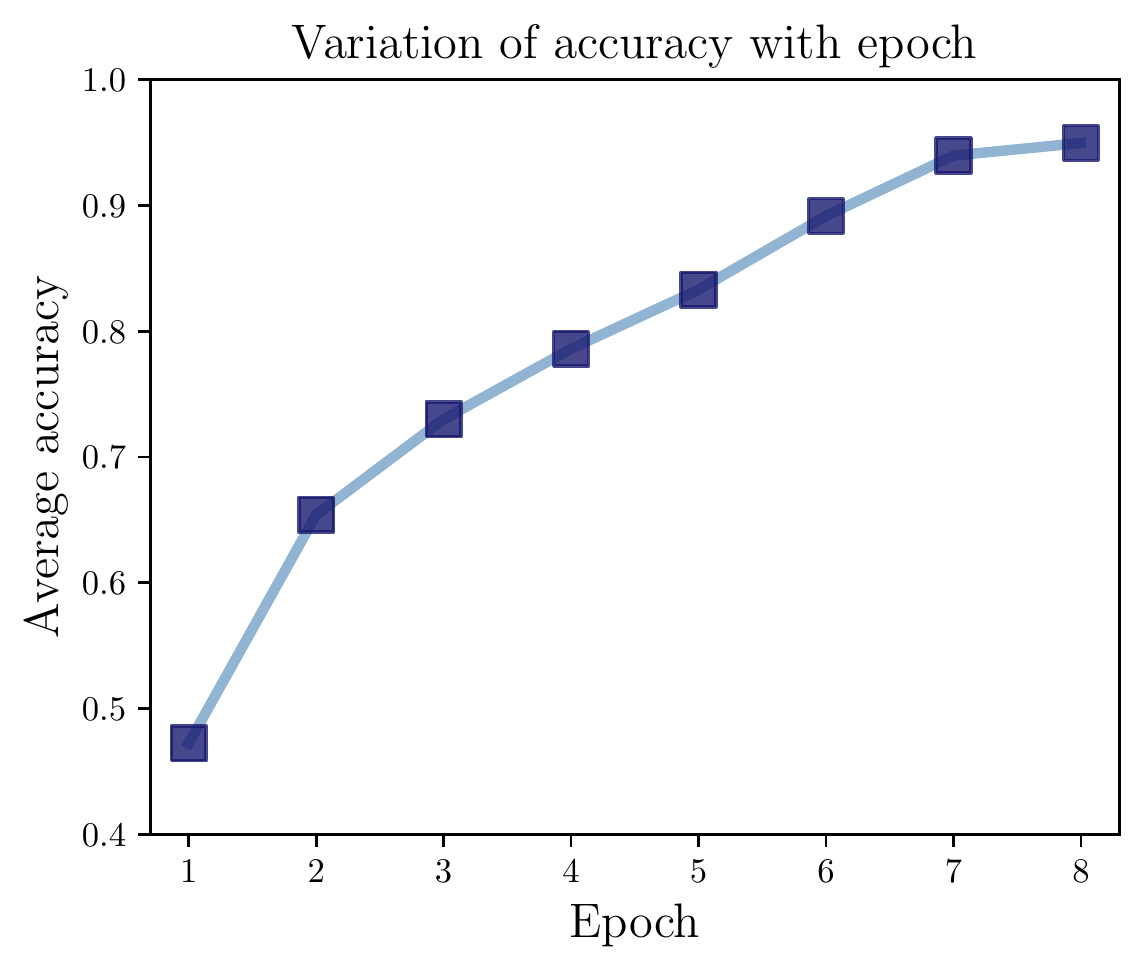}}
	\caption{Variation of the average classification accuracy of the spatio-temporal network with the number of observational epochs for the YSE set-up with an assumed cadence of six days. On average, the network is capable of identifying lensed sources with reasonable accuracy as from the third epoch itself. In practice, however, the potential for early detection is inevitably dependent on the seeing on the observing night and the characteristics of the lensed configuration (e.g. source redshift, position and Einstein radius).}
	\label{fig:av_accuracy_epoch}
\end{figure}

Early identification of lensed sources is crucial so as to rapidly follow up the lensed supernova for spectral characterisation purposes, and subsequently allow for opportunities for cosmological inference. In this context, correct lensed supernova identification using solely the first few epochs, without having to rely on all the epochs, is of paramount importance. To illustrate the potential for early detections, Fig.~\ref{fig:av_accuracy_epoch} indicates the average classification accuracy of the spatio-temporal network as a function of observational epochs for YSE-like lensed supernovae. This variation in accuracy shows that the network has the potential to detect lensed sources as from the third epoch, although, in practice, the early detection capabilities of our model for a given lensed supernova will primarily depend on the seeing on the observing night and the characteristics of the lensed configuration (in particular, source redshift, position and Einstein radius).

\begin{figure*}
	\centering
		{\includegraphics[width=\hsize,clip=true]{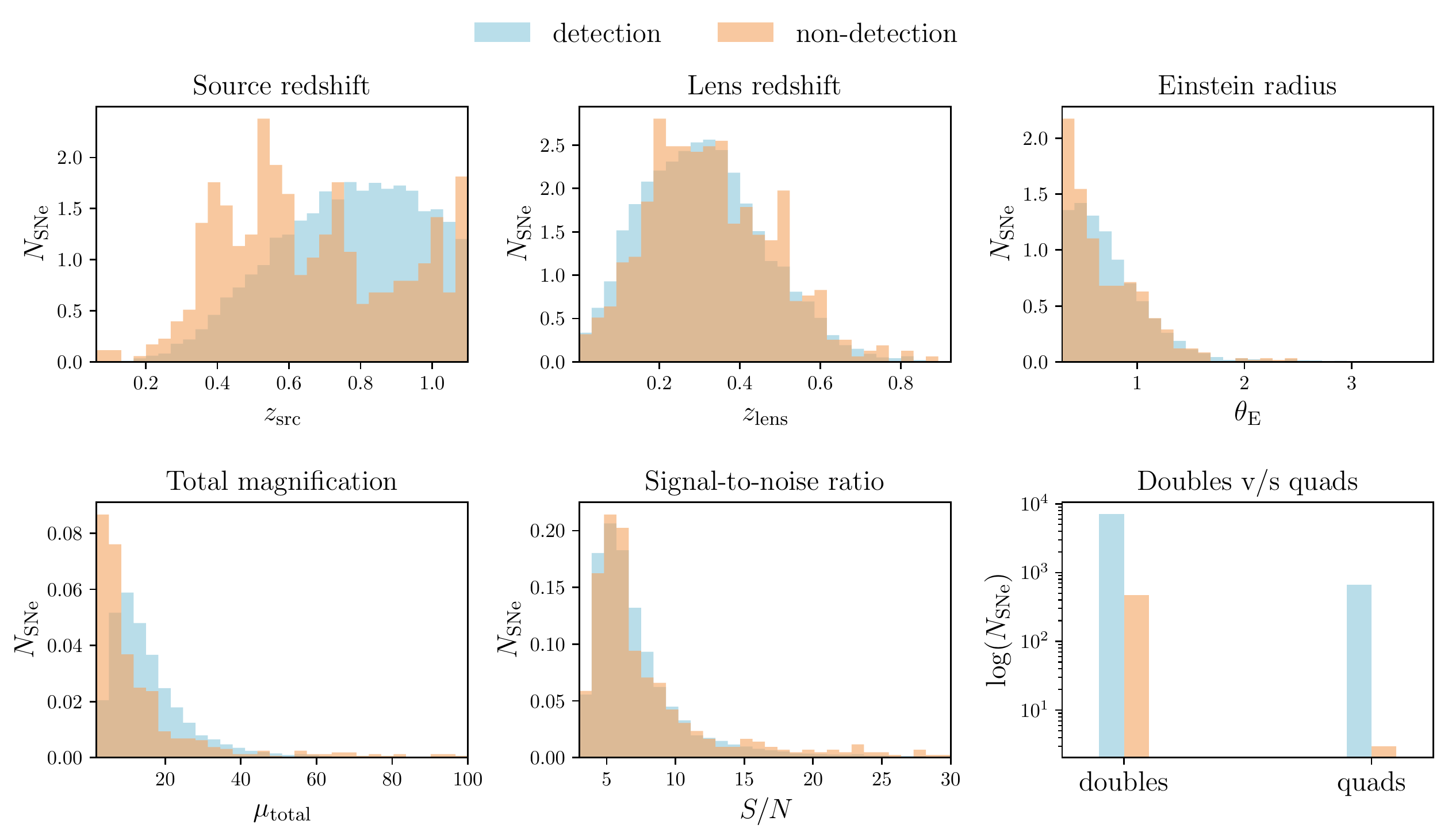}}
	\caption{Distributions of the properties of lensed supernovae, simulated within the YSE set-up, that are both correctly and wrongly classified by the spatio-temporal network. Note that the above histograms are normalised, with non-detections constituting only around 5\% of the test set, as indicated in Fig.~\ref{fig:confusion_matrix}{\color{RoyalBlue}(c)}, for the default ML classification threshold of 0.5. As expected, non-detections correspond primarily to systems with small Einstein radius ($\theta_{\mathrm{E}}$) and lower total magnification ($\mu_{\mathrm{total}}$). Quads are also more readily detected than doubles, with only three wrongly classified quads.}
	\label{fig:detection_vs_nondetection}
\end{figure*}

To obtain some insights related to the lensed supernovae that are both correctly and wrongly classified by the spatio-temporal network, the distributions of the relevant physical characteristics of the lensed systems are illustrated in Fig.~\ref{fig:detection_vs_nondetection}. We find that, as expected, misclassified lensed supernovae correspond to systems with small Einstein radii and lower values of total magnification. Undetected lensed supernovae at low redshifts are preferentially closer to their lens, which yields relatively small Einstein radii and in turn, typically lower angular image separations. The network is more proficient at finding quadruply imaged supernovae (i.e. quads) than their doubly imaged counterparts (i.e. doubles), which is also expected, given that quads are visually more distinctive and have relatively higher magnifications.

To demonstrate the capacity of our network at identifying real lensed supernova, we simulate the multi-epoch observations of the well-known lensed supernova iPTF16geu \citep{goobar2017iptf16geu} according to YSE settings using the known source and lens properties, $z_{\mathrm{src}} = 0.409, z_{\mathrm{lens}} = 0.21, \alpha = 0.0072^{\prime \prime}, \delta = 0.0012^{\prime \prime}, \theta_{\mathrm{E}} = 0.29^{\prime \prime}, \gamma_{\mathrm{lens}} = 2.0, q_{\mathrm{lens}} = 0.83$ \citep{more2017interpreting}, in accordance with the notation used in Section~\ref{methods}. We generate an ensemble of 100 realisations according to the YSE PSF distribution, including external shear and stochastic microlensing effects, that are subsequently fed to the trained spatio-temporal model. We find that the network identifies the simulated iPTF16geu supernova as a strong candidate with mean probability and confidence scores of 0.97 and 0.94, respectively.

Image subtraction artefacts, resulting from imperfect subtraction of the observed and reference images, are a common occurrence when dealing with difference images. Such artefacts are likely to be interpreted as potential lensing signatures by the neural classifier, thereby degrading its efficacy. To minimise the number of false positives due to such artefacts, we can compare the distribution of the apparent host-supernova separations and galaxy redshifts of the candidates from the spatio-temporal model with the distribution of Monte Carlo simulations of all detectable lensed type Ia supernovae simulated within a YSE set-up. We can subsequently discard the candidates that fall outside the region specified by the contours of the expected probability distribution of type Ia lensed supernovae in YSE, with such candidates likely to be supernovae in nearby galaxies whose large apparent sizes give rise to image subtraction artefacts.

\subsection{Application to mock LSST observations}
\label{application_mock_LSST}

Next-generation transient surveys are poised to drastically increase the number of known lensed supernovae. For instance, the Legacy Survey of Space and Time (LSST) from the Vera C.~Rubin Observatory \citep{ivezic2019lsst} should lead to the imminent discovery of several hundreds of lensed supernovae \citep{goldstein2017glSNe, wojtak2019magnified}, while the Nancy Grace Roman Space Telescope \citep{spergel2015roman} is expected to find a few dozens \citep{oguri2010gravitationally}. For such surveys with enhanced resolution and seeing, we expect an improved classification performance of the spatio-temporal model. To showcase the relevance of our spatio-temporal engine for such upcoming surveys, we verify the classification efficacy for mock observations generated within a preliminary LSST-like set-up (cf. Table~\ref{tab:YSE_specs}). The gold line in Fig.~\ref{fig:ROC_curve} depicts the resulting ROC curve with an AUC score of almost unity ($\mathrm{AUC} = 0.998$), with the corresponding confusion matrix provided in Fig.~\ref{fig:confusion_matrix} ($\mathrm{MCC} = 0.978$) indicating an overall classification accuracy of around 99 per cent. Over the lifetime of LSST, $\mathcal{O}(10^2)$ lensed supernovae are expected out of the $\mathcal{O}(10^5)$ new supernova discoveries per year \citep{goldstein2017glSNe, wojtak2019magnified}. With the spatio-temporal network, the odds of finding a lensed supernova rise from 1 in 1000 to 1 in 10, representing an improvement by two orders of magnitude.

\section{Conclusions and outlook}
\label{conclusions}

We presented a novel AI-assisted spatio-temporal engine, based on recurrent convolutional layers, to identify gravitationally lensed supernovae from the presently ongoing YSE survey. Our approach draws from recent advances in variational inference to quantify approximate Bayesian uncertainties, thereby assigning a confidence score to each model prediction that accurately reflects the uncertainty inherent to the network classification. Incorporating the temporal information encoded in the evolution of a given source led to a significant gain of around 20 per cent in classification accuracy relative to single-epoch observations for the test data set generated within the YSE set-up. Such neural classifiers are complementary to the standard selection based on typical lens galaxy redshifts and lens-supernova angular separations, with the combination of these two distinct approaches crucial for rapidly identifying promising lensed supernova candidates, such that follow-up spectroscopic observations can be initiated in a timely manner.

Our spatio-temporal model is tailored to detect time-variable lensing features in the time series of difference images. To illustrate that the temporal correlations in the time-series images of a given source are conducive to the classification accuracy, we trained a CNN using the compressed time-series representation. To this end, we implemented a variant of the smooth manifold extraction technique, originally proposed for processing a stack of images produced via 3D fluorescence microscopy, to compress the observed time-series images of an astrophysical transient into a single informative image. We find that the spatio-temporal model results yet in an improvement of over 10 per cent in accuracy relative to its compressed temporal counterpart. This gain in accuracy matches our intuitive expectations since the variation in observed brightness of lensed supernovae has a particular trend.

Having an estimate of the confidence associated with a given neural classifier prediction undeniably brings some additional insights when sifting through an avalanche of plausible transients as recorded by an instrument. Human vetting, as the typical final step after the ML-based pre-filtering, should benefit from the extra information about the neural network uncertainties when assessing and prioritising potential lensed supernovae. Moreover, the confidence metric can also be incorporated alongside the network probability score on the data acquisition platform for prompt detection alerts concerning extremely promising lensed candidates. This opens the possibility of automating decision-making pertaining to follow-up observations and reporting.

The AI machinery presented here may be naturally adapted in the context of upcoming transient surveys that would deliver unprecedented volumes of lensed candidates. Indeed, we performed a preliminary application to LSST-like difference images and showed that the classification accuracy of the spatio-temporal model rises to around 99 per cent, thereby demonstrating its efficacy for next-generation surveys. The classification accuracy can be even further enhanced when images in the remaining filters are included, thereby improving an effective temporal resolution of the image series. As such, a straightforward extension is to include additional channels in our network so as to work at the level of multi-band images. Moreover, to render our framework more robust to both lensed supernova impostors (e.g. AGN) and image subtraction artefacts, a plausible extension is to incorporate real YSE observations in the training set.

An interesting application of our spatio-temporal engine is to determine gravitational time-delays from multiple-epoch images of lensed supernovae and subsequently infer the Hubble constant \citep{arendse2021h0}, which constitutes our ongoing study. This yields a complementary framework to standard cosmographic analyses in the quest for independent measurements of the cosmic acceleration. Other potential applications of our spatio-temporal framework include supernova spectral classification from multiple-epoch optical images, identification of reionization sources from 21 cm lightcones, distinguishing dark energy models and identifying structures of the cosmic web from cosmic structure lightcones, highlighting the broader applicability of our ML framework.

\section*{Acknowledgements}

We thank Jens Hjorth, David Jones and Ken Smith for their assistance with our YSE-related queries. We thank Tom Charnock for his views on variational inference and David Prelogovi\'c for interesting exchanges about recurrent networks. DKR is a DARK fellow supported by a Semper Ardens grant from the Carlsberg Foundation (reference CF15-0384). This work was supported by a VILLUM FONDEN Investigator grant (project number 16599). This work has made use of the Infinity Cluster hosted by Institut d'Astrophysique de Paris and an HPC facility funded by a grant from VILLUM FONDEN (project number 16599).

\section*{Data availability}

The source code repository and catalogue of simulated images used in our analysis will be made available at \url{https://github.com/doogesh/spatio_temporal_glSNe_finder} upon publication.

\section*{Author contributions}

DKR led the project, built the image generation pipeline, implemented a variant of the SME algorithm for combining time-series snapshots, designed and optimised the neural network architecture (three ML models), performed the analysis, produced the figures, wrote the paper. NA implemented the code for generating time-series images, visualised the joint distribution of the lens properties, assisted the network optimisation and contributed to writing-up the manuscript. RW proposed the main idea, validated the image simulation pipeline, and contributed to writing-up the manuscript.




\bibliographystyle{mnras}
\bibliography{./compiled_references} 




\bsp	
\label{lastpage}
\end{document}